\documentclass{pasa}%

\title[The ASKAP/EMU Data Challenge]{The ASKAP/EMU Source Finding Data Challenge}
\author[A. M. Hopkins et al.]{A. M. Hopkins$^{1,}$\thanks{Email: ahopkins@aao.gov.au}, M. T. Whiting$^{2}$,
N. Seymour$^{3}$, K. E. Chow$^{2}$, R. P. Norris$^{2}$, L. Bonavera$^{4}$, R. Breton$^{5}$, D. Carbone$^{6}$,
C. Ferrari$^{7}$, T. M. O. Franzen$^{3}$, H. Garsden$^{8}$, J. Gonz{\'a}lez-Nuevo$^{9,4}$, C. A. Hales$^{10,11}$,
P. J. Hancock$^{3,12,13}$, G. Heald$^{14,15}$, D. Herranz$^{4}$, M. Huynh$^{16}$, R. J. Jurek$^{2}$,
M. L{\'o}pez-Caniego$^{17,4}$, M. Massardi$^{18}$, N. Mohan$^{19}$, S. Molinari$^{20}$, E. Orr{\`u}$^{14}$,
R. Paladino$^{21,18}$, M. Pestalozzi$^{20}$, R. Pizzo$^{14}$, D. Rafferty$^{22}$, H. J. A. R{\"o}ttgering$^{23}$,
L. Rudnick$^{24}$, E. Schisano$^{20}$, A. Shulevski$^{14,15}$, J. Swinbank$^{25,6}$, R. Taylor$^{26,27}$,
A. J. van der Horst$^{28,6}$\\
\affil{$^1$ Australian Astronomical Observatory, PO Box 915, North Ryde, NSW, 1670, Australia}
\affil{$^2$ CSIRO Astronomy \& Space Science, PO Box 76, Epping, NSW 1710, Australia}
\affil{$^3$ International Centre for Radio Astronomy Research, Curtin University, GPO Box U1987, Perth WA 6845, Australia}
\affil{$^4$ Instituto de F{\'i}sica de Cantabria (CSIC-UC), Santander, 39005 Spain}
\affil{$^5$ Jodrell Bank Centre for Astrophysics, The University of Manchester, Manchester, M13 9PL, UK }
\affil{$^6$ Anton Pannekoek Institute for Astronomy, University of Amsterdam, Postbus 94249, 1090 GE Amsterdam, The Netherlands }
\affil{$^7$ Laboratoire Lagrange, Universit{\'e} C{\^o}te d'Azur, Observatoire de la C{\^o}te d'Azur, CNRS,
Blvd de l'Observatoire, CS 34229, 06304 Nice cedex 4, France}
\affil{$^8$ Laboratoire AIM (UMR 7158), CEA/DSM-CNRS-Universit{\'e} Paris Diderot, IRFU, SEDI-SAP, Service dÕAstrophysique, Centre de Saclay, F-91191 Gif-Sur-Yvette cedex, France}
\affil{$^9$ Departamento de F\'{i}sica, Universidad de Oviedo, C. Calvo Sotelo s/n, 33007 Oviedo, Spain}
\affil{$^{10}$ National Radio Astronomical Observatory, P.O. Box O, 1003 Lopezville Road, Socorro, NM 87801-0387, USA}
\affil{$^{11}$ Jansky Fellow, National Radio Astronomical Observatory}
\affil{$^{12}$ Sydney Institute for Astronomy, School of Physics A29, The University of Sydney, NSW 2006, Australia}
\affil{$^{13}$ ARC Centre of Excellence for All-Sky Astrophysics (CAASTRO)}
\affil{$^{14}$ ASTRON, the Netherlands Institute for Radio Astronomy, Postbus 2, 7990 AA, Dwingeloo, The Netherlands}
\affil{$^{15}$ University of Groningen, Kapteyn Astronomical Institute, Landleven 12, 9747 AD Groningen, The Netherlands}\affil{$^{16}$ International Centre for Radio Astronomy Research, M468, University of Western Australia, Crawley, WA 6009, Australia}
\affil{$^{17}$ European Space Agency, ESAC, Planck Science Office, Camino bajo del Castillo, s/n, Urbanizaci\'{o}n
Villafranca del Castillo, Villanueva de la Ca\~{n}ada, Madrid, Spain}
\affil{$^{18}$ INAF-Istituto di Radioastronomia, via Gobetti 101, 40129 Bologna, Italy}
\affil{$^{19}$ National Centre for Radio Astrophysics, Tata Institute of Fundamental Research, Post Bag 3, Ganeshkhind, Pune 411 007, India}
\affil{$^{20}$ IAPS - INAF, via del  Fosso del Cavaliere 100, I - 00173 Roma, Italy}
\affil{$^{21}$ Department of Physics and Astronomy, University of Bologna, V.le Berti Pichat 6/2, 40127 Bologna, Italy}
\affil{$^{22}$ Hamburger Sternwarte, Universit{\"a}t Hamburg, Gojenbergsweg 112, D-21029 Hamburg, Germany}
\affil{$^{23}$ Leiden Observatory, Leiden University, P.O. Box 9513, 2300 RA, The Netherlands}
\affil{$^{24}$ Minnesota Institute for Astrophysics, University of Minnesota, 116 Church St. SE, Minneapolis, MN 55455}
\affil{$^{25}$ Department of Astrophysical Sciences, Princeton University, Princeton, NJ 08544, USA}
\affil{$^{26}$ Department of Astronomy, University of Cape Town, Private Bag X3, Rondebosch, 7701, South Africa}
\affil{$^{27}$ Department of Physics, University of the Western Cape, Robert Sobukwe Road, Bellville, 7535, South Africa}
\affil{$^{28}$ Department of Physics, The George Washington University, 725 21st Street NW, Washington, DC 20052, USA}
}%
\jid{PASA}
\doi{10.1017/pas.\the\year.xxx}
\jyear{\the\year}

\usepackage[authoryear]{natbib}
\bibpunct{(}{)}{;}{a}{}{,}
\usepackage{graphicx}

\begin{document}%
\begin{abstract}
The Evolutionary Map of the Universe (EMU) is a proposed radio continuum survey of the Southern
Hemisphere up to declination $+30^{\circ}$, with the Australian Square Kilometre Array Pathfinder
(ASKAP). EMU will use an automated source identification and measurement approach that is
demonstrably optimal, to maximise the reliability, utility and robustness of the resulting radio source catalogues.
As part of the process of achieving this aim, a ``Data Challenge" has been conducted, providing international
teams the opportunity to test a variety of source finders on a set of simulated images. The aim is to quantify the
accuracy of existing automated source finding and measurement approaches, and to identify potential
limitations. The Challenge attracted nine independent teams, who tested eleven different source finding tools.
In addition, the Challenge initiators also tested the current ASKAPsoft source-finding tool to establish
how it could benefit from incorporating successful features of the other tools.
Here we present the results of the Data Challenge, identifying the successes and limitations
for this broad variety of the current generation of radio source finding tools. As expected, most finders
demonstrate completeness levels close to 100\% at $\approx 10\,\sigma$ dropping to levels around
10\% by $\approx 5\,\sigma$. The reliability is typically close to 100\% at $\approx 10\,\sigma$, with
performance to lower sensitivities varying greatly between finders. All finders demonstrate the
usual trade-off between completeness and reliability, whereby maintaining a high completeness at
low signal-to-noise comes at the expense of reduced reliability, and vice-versa.
We conclude with a series of recommendations for improving the performance of the ASKAPsoft source-finding tool.
\end{abstract}
\begin{keywords}
methods: data analysis --- radio continuum: general --- techniques: image processing
\end{keywords}
\maketitle%
\section{INTRODUCTION }
\label{sec:intro}

Measuring the properties of astronomical sources in images produced by radio interferometers
has been successfully achieved for many decades through a variety of techniques. Probably the
most common in recent years has been through identifying local peaks of emission above some
threshold, and fitting two-dimen\-sional Gaussians \citep[e.g.,][]{Con:97}. This approach is in principle substantially
unchanged from the very earliest generation of automated source detection and measurement approaches
in radio interferometric imaging. These also used a thresholding step followed by integration of the
flux density in peaks of emission above that threshold \citep[e.g.,][]{Ken:66}. This in turn
followed naturally from the earlier practice of defining a smooth curve through the
minima of paper trace profiles to represent the background level \citep[e.g.,][]{Lar:61}.

A variety of automated tools for implementing
this approach have been developed. In almost all cases the automatically determined source
list requires some level of subsequent manual adjustment to eliminate spurious detections
or to include objects deemed to be real but that were overlooked by the automated finder. This
manual adjustment step, again, has remained unchanged since the earliest days of radio source
measurement \citep[e.g.,][]{HM:62}.

As radio surveys have become deeper and wider, and the numbers of sources in the automated
catalogues becomes large, such manual intervention is progressively less feasible.
The FIRST survey \citep{Whi:97} contains about 900\,000 sources, and the NVSS \citep{Con:98}
about 1.8 million sources. In the case of future wide-area and deep surveys with new telescope
facilities, such as the Australian Square Kilometre Array Pathfinder \citep[ASKAP,][]{Jon:07},
this number will be increased by substantially more than an order of magnitude. The
Evolutionary Map of the Universe \citep[EMU,][]{Nor:11}, for example, is expected to yield
about 70 million radio sources. The full Square Kilometre Array will produce orders of magnitude
more again \citep[e.g.,][]{Hop:00}.

There has been a strong movement in recent years to ensure that the automated source detection
pipelines implemented for these next generation facilities produce catalogues with a high degree
of completeness and reliability, together with well-defined and characterised measurement accuracy.
Several recent analyses explore the properties of various source-finders, and propose refinements
or developments to such tools \citep[e.g.,][]{Pop:12,Huy:12,Hal:12,Han:12,Moo:13,Per:15}.
At the second annual SKA Pathfinder Radio Continuum Surveys (SPARCS) workshop, held in Sydney over
2012 May 30 to 2012 June 1, many of these results were presented and discussed. A consensus was
reached that a blind source finding challenge would be a valuable addition to our current approaches for
understanding the strengths and limitations of the many source-finding tools and techniques
presently available. The Data Challenge presented here was initiated as a result. The intended
audience for this work includes not only the ASKAP team working on source finding solutions, but
also the developers of astronomical source finding and related tools, and potential coordinators of
future Data Challenges. The outcomes of this work have applicability to all these areas.

The goal of the Data Challenge is to assess the completeness, reliability, accuracy, and
common failure modes, for a variety of source-finding tools. These statistics and outcomes are presented
below for all the tools tested in the Challenge. The outcomes are being used to directly inform developments within
the ASKAP source finding pipeline. The primary focus is on ensuring that the ASKAP source finder is as robust
as possible for producing the EMU source catalogue, although these results are clearly of broad utility, in
particular for many of the current SKA Pathfinders and surveys.

The scope of the current Challenge is limited intentionally to point-sources or point-like sources (sources only
marginally extended), due to the inherent difficulty faced by automated source finders in dealing with complex
source structure. We do test the performance of such finders on somewhat extended sources in our
analysis, although given this limitation we do not explore such performance in great detail. This is clearly an area
that deserves more explicit attention, with a focus on how to develop automated source finders that accurately
characterise extended source structure \citep[e.g.,][]{HJH:12,Fre:14}.
Even with this limitation, there is clearly still much that can be learned about the approach
to automating a highly complete and reliable point source detection tool. It is hoped that future Data Challenges
will follow from this initial effort, exploring more complex source structures, as well as innovative approaches
to the source detection and characterisation problem.

Below we describe the Data Challenge itself (\S\,\ref{dc}) and the construction of the artificial images
used (\S\,\ref{ai}). This is followed by our analysis of the performance of the submitted finders (\S\,\ref{analysis})
and a discussion comparing these results (\S\,\ref{disc}). We conclude in \S\,\ref{conc} with a summary of the
outcomes.

\section{THE DATA CHALLENGE}
\label{dc}

The Data Challenge originators (Hopkins, Whiting and Seymour) had responsibility for preparing
the artificial source lists and images for the Challenge, initiating and promoting it to potential participants
and coordinating the Challenge itself, as well as the primary analysis of the outcomes.
The Data Challenge required teams to register their participation by 2012 November 30.
Three artificial images were provided, along with a selection of ancillary data detailed below.
The deadline for submitting the three source lists for each registered source finder was 2013 January 15.

Participating teams were instructed to provide details of the source finding tool being tested, including the
name, version number if appropriate, instructions for obtaining the tool itself, and any other information to uniquely
identify the tool and mode or parameters of operation as relevant. The teams were also required to identify any
potential conflicts of interest that may have biased or influenced their analysis, or prevented the analysis from
being truly blind. No such conflicts were identified by any participating teams.

The source lists submitted by the teams were required to have file names allowing them to be uniquely
associated with the corresponding Challenge image. The format of each source list was required
to be a simple ascii text file containing one line per source, with a header line (or lines) marked by a hash (\#)
as the initial character, to uniquely define the columns of the ascii table. The columns were required to
include RA and Dec, peak and integrated flux density, deconvolved semi-major axis, deconvolved semi-minor axis, and
position angle. Errors on all of these quantities were also requested.
Multiple submissions were acceptable if teams desired to have different operating modes or parameter
sets for a given tool included in the analysis. Several of the submitted finders included multiple different modes
or parameter settings, and these are referred to in the text and figures below by the name of the finder followed
by the mode of use in brackets. Not all submissions included size and position angle measurements,
as not all tools tested necessarily provide those measurements. The list of tools submitted, with published
references to the tool where available, is given in Table~\ref{sftools}. A brief description of each finder and how it
was used in the Challenge is presented in Appendix~\ref{app1}. We note that some finders may need
considerable fine tuning of parameters and consequently the conclusions presented from this Challenge
reflect the particular finder implementation used for these tests.

In addition to the tools submitted by participants, two additional tools were tested by the Challenge originators.
These are Duchamp and Selavy, tools that were both authored by Whiting, and multiple modes for these finders
were tested. While all care has been taken
to treat these tools and their outputs objectively, we acknowledge the conflict of interest present, and these
cannot be assessed in a truly ``blind'' fashion as with the other tools tested. Bearing this in mind, we felt
that it would be valuable to identify the strengths and weaknesses of these tools in the same way as
the others that are being tested in a truly ``blind'' fashion. The intent is to identify elements of the best-performing
tools that can subsequently be incorporated into the ASKAP source finder, or common failure modes
that can be eliminated if present. Note that Selavy is the current prototype for the ASKAP pipeline-based source-finder.

\begin{table*}[h]
\caption{List of source finding tools tested}
\begin{center}
\begin{tabular}{@{}llll@{}}
\hline\hline
Source finder & Submitter or Team & Reference  \\
\hline
Aegean & P.\ Hancock & \citet{Han:12} \\
APEX & M.\ Huynh & \\
{\sc blobcat} & C.\ Hales & \citet{Hal:12} \\
CuTEx & IAPS-INAF & \citet{Mol:11} \\
IFCA BAF & IFCA & \citet{Lop:12} \\
IFCA MF & IFCA & \citet{Lop:06} \\
PyBDSM & LOFAR & \citet{MR:15} \\
PySE & LOFAR & \citet{Swi:15,Spr:10} \\
SAD & L.\ Rudnick \& R.\ Taylor &  \\
SExtractor & M.\ Huynh & \citet{Ber:96}  \\
SOURCE\_FIND & T.\ Franzen & \citet{Fra:11} \\
\hline
Duchamp & M.\ Whiting & \citet{Whi:12}  \\
Selavy & M.\ Whiting & \citet{WH:12}  \\
\hline\hline
\end{tabular}
\end{center}
\label{sftools}
\end{table*}

\begin{figure*}[ht]
\begin{center}
\includegraphics[width=15cm, angle=0]{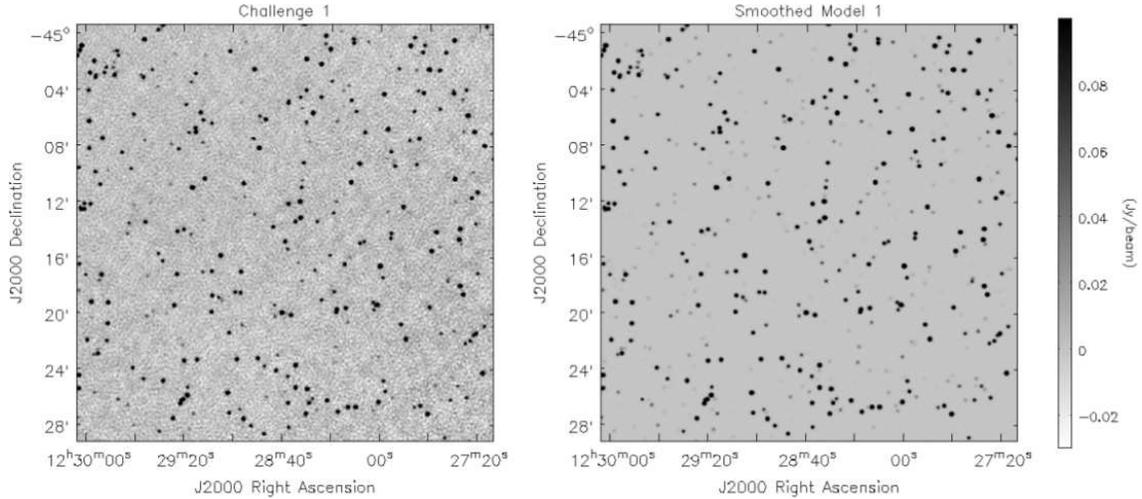}
\caption{A subsection of the first Data Challenge image (left), and the input source distribution to this image (right).
This image includes sources distributed randomly with a flux density distribution that is uniform in the logarithm of
flux density. This distribution gives rise to a much higher surface density of bright sources, and proportionally more
bright sources compared to faint sources, than in the real sky.}\label{challenge1}
\end{center}
\end{figure*}

\begin{figure*}[ht]
\begin{center}
\includegraphics[width=15cm, angle=0]{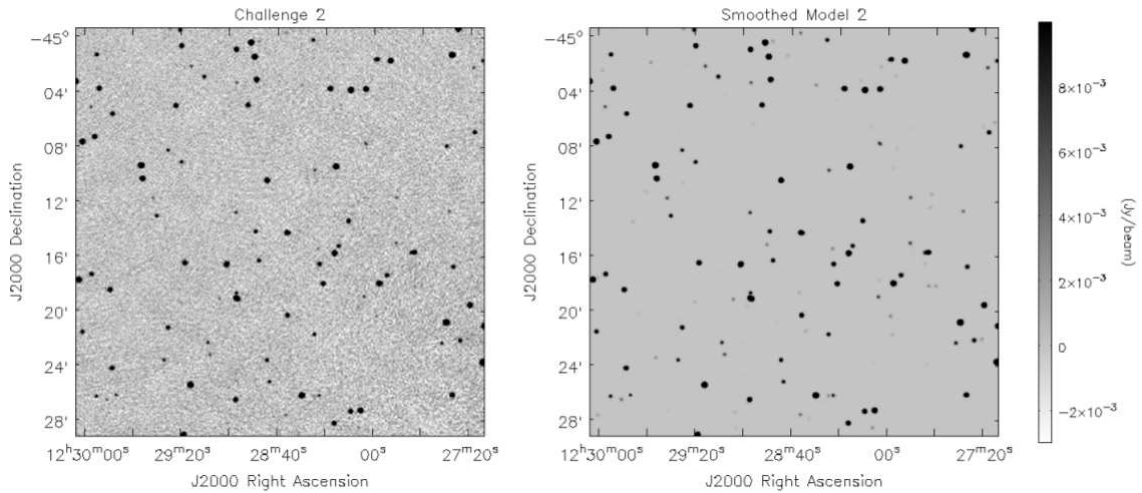}
\caption{A subsection of the second Data Challenge image (left), and the input source distribution to this image (right).
This image includes sources distributed with intrinsic clustering, and with a flux density distribution
drawn from the observed source counts \citep[e.g.,][]{Hop:03}, in an effort to mimic the characteristics of the real sky.}\label{challenge2}
\end{center}
\end{figure*}

\begin{figure*}[ht]
\begin{center}
\includegraphics[width=15cm, angle=0]{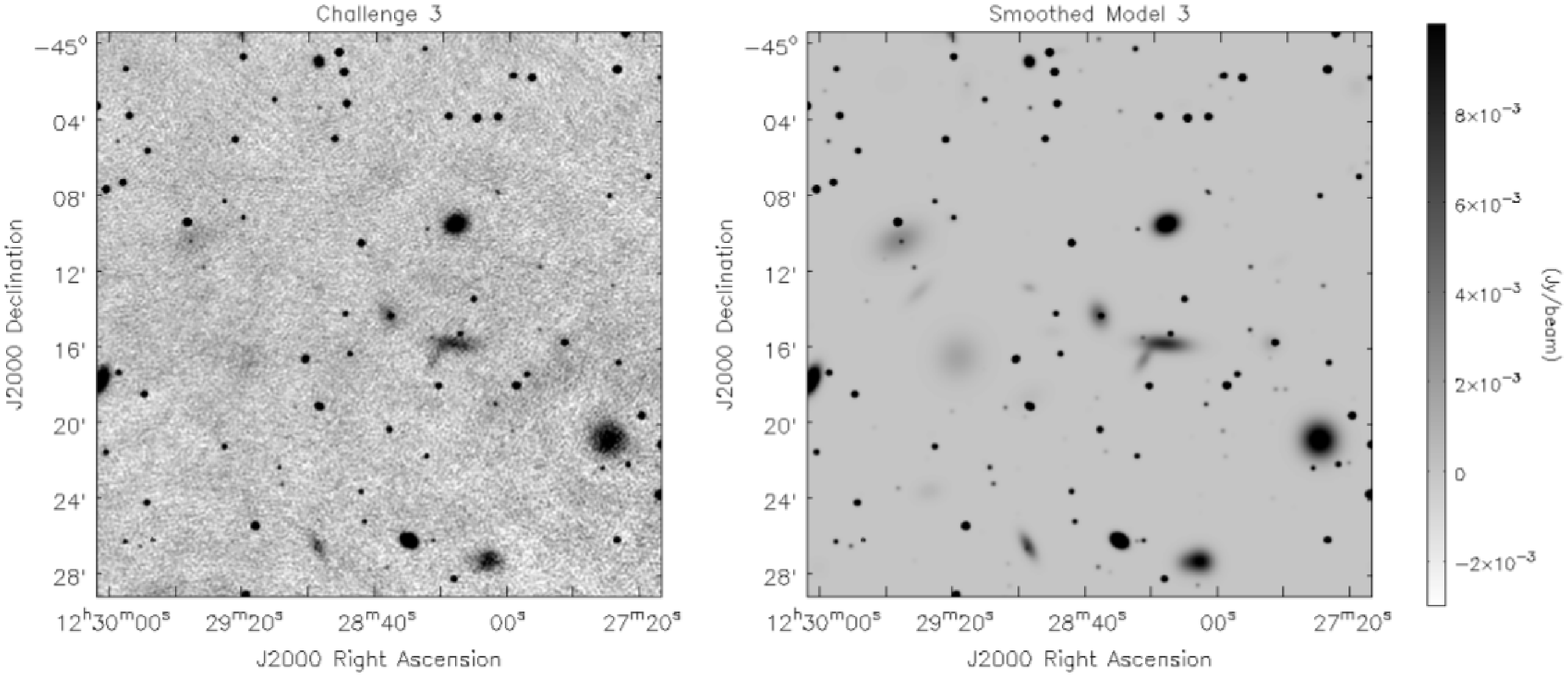}
\caption{A subsection of the third Data Challenge image (left), and the input source distribution to this image (right).
This image includes sources as for the second Data Challenge image, but with $20\%$ of the
sources now assigned a non-negligible physical extent. The extended sources are modelled as
two-dimensional elliptical Gaussians.}\label{challenge3}
\end{center}
\end{figure*}

\section{ARTIFICIAL IMAGE CREATION}
\label{ai}

\subsection{Artificial source catalogues}

For the Data Challenge images we created three input catalogues:

\begin{enumerate}
\item A bright source catalogue (Figure~\ref{challenge1}). The purpose of this test was to obtain an initial
    comparison of the different methods and to search for subtle systematic effects. We were interested in
    assessing the performance of source finders in a non-physical scenario to aid in determining whether there were
    any aspects of the real sky that influenced the outcomes in subtle ways.
    We created a catalogue with a surface density of about 3800 sources per square degree (the image synthesised beam
    size is $\approx 11''$, details in \S\,\ref{sec-images})
    with a uniform distribution in logarithmic flux density spanning
    $0.2<S_{1.4GHz}$(mJy)$<1000$. The positions were randomly assigned, with RA and Dec values for
    each source having a uniform chance of falling anywhere within the field.
\item A fainter catalogue with pseudo-realistic clustering and source counts (Figure~\ref{challenge2}). 
    Here we used a surface density of about 800 sources per square degree (the image synthesised beam size is
    $\approx 11''$, details in \S\,\ref{sec-images}) and had an increasing number of faint sources as measured in bins
    of logarithmic flux density, to mimic the real source counts \citep[e.g.,][]{Hop:03,Nor:11}. Sources were
    assigned flux densities in the range $0.04<S_{1.4GHz}$(mJy)$<1000$. The distribution of source
    positions was designed to roughly correspond to the clustering distributions measured by \citet{BW:02} for sources having
    $S_{1.4GHz}>1\,$mJy, and to \citet{Oli:04} for $S_{1.4GHz}<1\,$mJy. In the latter case we assume that
    faint radio sources have similar clustering to faint IRAC sources, in the absence of explicit clustering measurements
    for the faint population, and on the basis that both predominantly reflect a changing proportion of low
    luminosity AGN and star forming galaxy populations. In each case we began with
    an initial random list of source locations, then used an iterative process to test the clustering signal in the vicinity of
    each source, relocating neighbour sources until the desired clustering was reproduced.
\item The same as (2), but with some sources extended (Figure~\ref{challenge3}). We randomly designated 
    $20\%$ of those sources to be elliptical Gaussians with the total 
    flux conserved (and therefore having a lower peak flux density). These elliptical Gaussians
    were assigned major axis lengths of $5''$ to $140''$, with 
    brighter sources likely to be more extended than fainter ones. 
    The minor axis length was then randomly varied between $30\%$ and $100\%$ 
    of the major axis length. 
\end{enumerate}

\begin{figure}[ht]
\begin{center}
\includegraphics[width=7cm, angle=0]{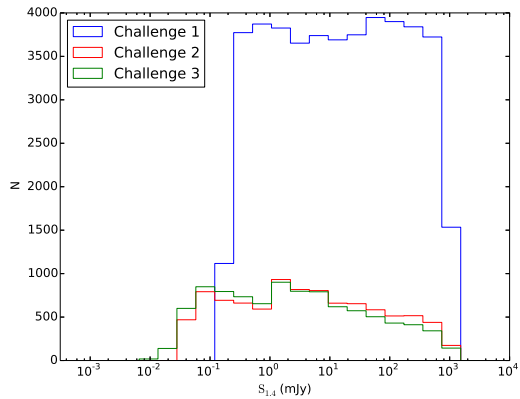}
\caption{The distribution of input source flux densities for the three Challenges.}\label{challenge_counts}
\end{center}
\end{figure}

In Figures~\ref{challenge1}-\ref{challenge3} we show subsections of the Challenge images used and the
input source models, in order to illustrate cleanly the characteristics of the noise in the images.
The input sources were assigned to have flat spectra ($\alpha=0$,
where $S_\nu \propto \nu^{\alpha}$) as investigations of spectral index 
effects are beyond the scope of these tests. For all three Challenges, the distribution of
source flux densities end up spanning a broad range of signal-to-noise (S/N), from S/N$<1$
to S/N$>100$ (Figure~\ref{challenge_counts}). Each catalogue covered a
square region of $30\,\deg^2$ (to match the ASKAP field-of-view) and
were centred arbitrarily at RA$=12^h30^m$, Dec$=-45^{\circ}$.

\subsection{Artificial image generation}
\label{sec-images}

The images were created with two arcsecond pixels. To simplify the
computational elements of the imaging each source was shifted slightly to be at the 
centre of a pixel. If multiple sources were shifted to a common location they were simply
combined into a single input source by summing their flux densities. This step had a negligible
effect on both the implied clustering of the sources and the input flux density distribution,
but a significant reduction in the computational requirements for producing the artificial images.
The input source catalogue used subsequently to assess the performance of the submitted finders
was that produced after these location shifting and (if needed) flux combining steps.
Simulating a more realistic distribution of source positions should be explored in future work
to assess the effect on finder performance for sources lying at pixel corners rather than pixel centres.

The image creation step involves mimicking the process of observation, populating the
{\em uv\/} plane by sampling the artificial noise-free sky for a simulated 12 hour synthesis with
the nominal 36 ASKAP antennas, adding realistic noise (assuming $T_\text{sys}=50K$ and
aperture efficiency $\eta=0.8$) to the visibilities. Noise was added in the {\em uv\/} plane in
the XX and YY polarisations with no cross-polarisation terms. This simulates the thermal noise in
the visibilities in order to correctly mimic the behaviour of the real telescope. The image-plane
noise consequently incorporates the expected correlation over the scale of the restoring beam.
Because of a limitation in computing resources, a reduced image size compared to that
produced by ASKAP was simulated giving a field of view of $15.8\,\deg^2$ (or $11.6\,\deg^2$
once cropped, described further below), as it was judged this was sufficient to provide a large
number of sources yet still keep the images of a manageable size for processing purposes. 
The visibilities were then imaged via Fourier transformation and deconvolution.
The deconvolution step was run for a fixed number of iterations for each of the three Challenge images.
As a consequence of this step, limited by available CPU time for this compute-intensive process,
the image noise level in the simulations is significantly higher than the nominal theoretical noise.
This is exacerbated by the presence of many faint sources below the measured noise level in the
simulated images. We emphasise that the processing of real ASKAP images will not be limited in this way.
For Challenge~1 the noise level was higher, by almost a factor of 10, than in the images for Challenges~2 and 3.
We attribute this, and the subsequent low dynamic range in the flux-density distribution of sources able to be measured
in Challenge 1, to the non-physical distribution of flux densities resulting from the high surface density of bright sources.

Due to the density of sources on the sky, especially for Challenge~1, and with
clustering (random or not) many sources were close enough together that they were either
assigned to the same pixel, or would fall within the final restored beam of the image
($11.2''\times10.7''$, PA$=3.1^\circ$) of an adjacent source.
While sources with their peaks lying within the same resolution element may be able to be
distinguished, given sufficient S/N depending on the separation, the bulk
of measured radio sources in large surveys are at low S/N. Even sources in this regime with their peaks
separated by more than one resolution element but still close enough to overlap are clearly a
challenge \citep{Han:12}, even without adding the extra complexity of
sources lying within a common resolution element. To avoid making the Data Challenge
too sophisticated initially and to focus on the most common issues, for Challenges~1 and 2 all sources from the
preliminary input catalogue that lay within $11''$ of each other were replaced by a single source
defined as the combination of the preliminary sources by adding their fluxes
and using the flux weighted mean positions. While most of these matches were
pairs we also accounted for the small number of multiple such matches.

For Challenge~3 with $20\%$ of the sources potentially quite extended we had 
to employ a different method. For relatively compact sources, defined as those
having a major axis $<16.8''$ ($1.5\times$FWHM), we combined them as before if
they were isolated from extended sources. For the rest of the sources with larger
extent we simply flagged them as either being isolated if no other sources overlapped
the elliptical Gaussian, or as being blended. 

For comparison with the submitted catalogues, we restricted both the simulated and
measured catalogues to areas that had good sensitivity, removing the edges of the image
where the image noise increased. In practice, we applied a cutoff where the theoretical
noise increased by a factor of 2.35 over the best (lowest) noise level in the field.

\section{ANALYSIS}
\label{analysis}

\subsection{Completeness and reliability}
\label{CandR}

\begin{figure*}[ht]
\begin{center}
\centerline{\includegraphics[width=9cm]{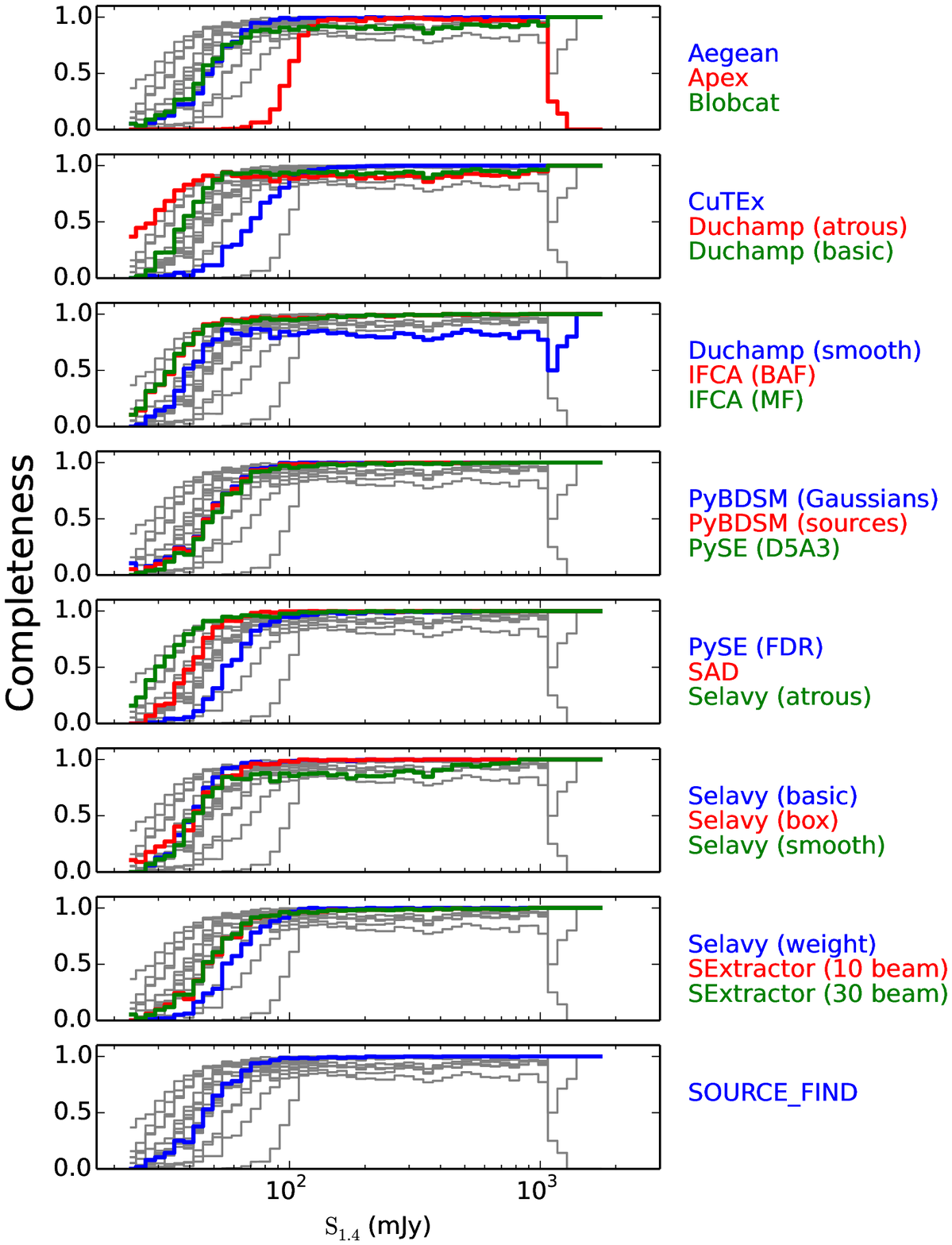}\hspace{-2mm}
\includegraphics[width=9cm]{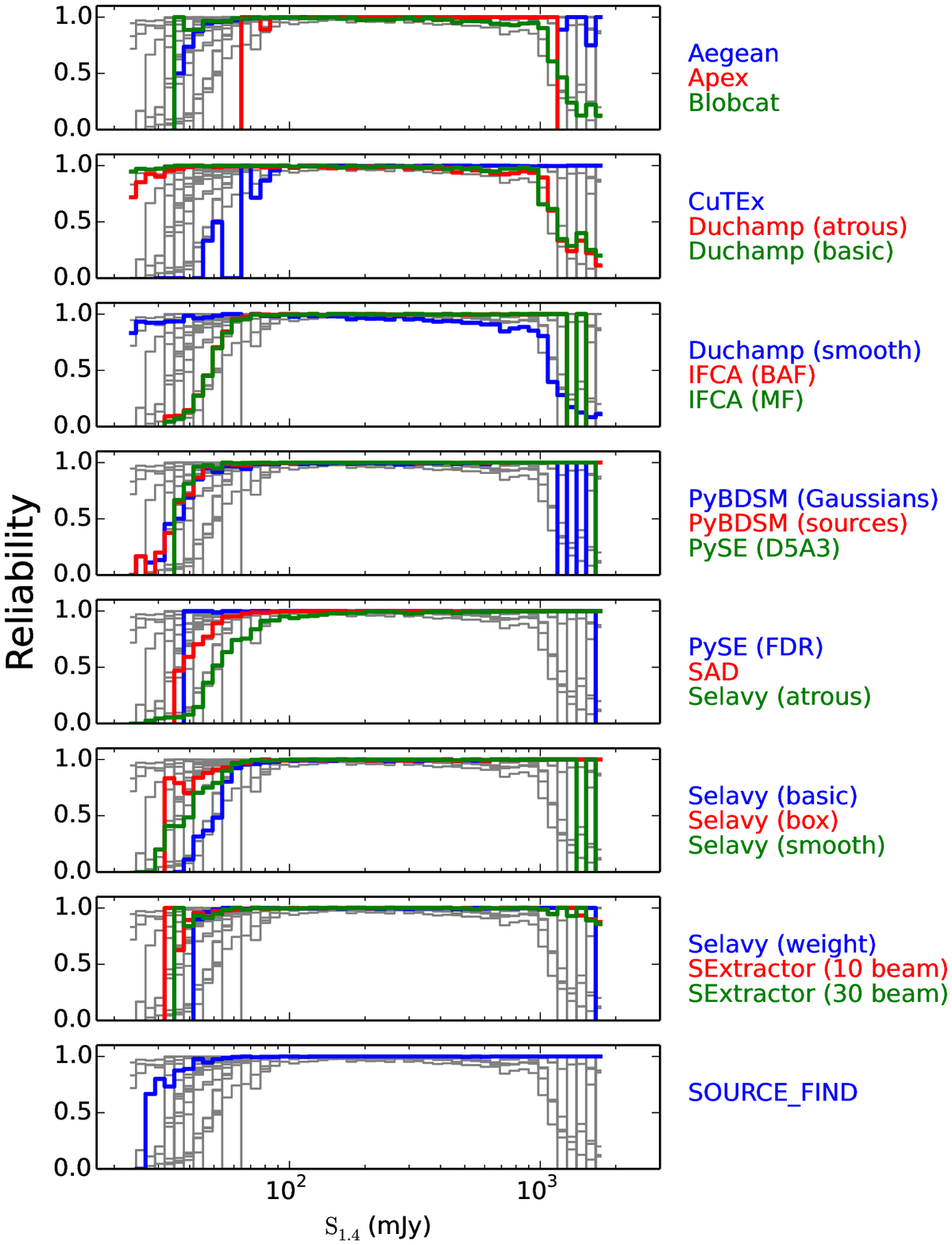}}
\caption{The completeness and reliability fractions (left and right respectively) as a function of
input source flux density (completeness) or measured source flux density (reliability) for each of the tested
source finders for Challenge~1. The grey lines
show the distribution for all finders in each panel, to aid comparison for any given finder.}\label{completeness1}
\end{center}
\end{figure*}

\begin{figure*}[ht]
\begin{center}
\centerline{\includegraphics[width=9cm]{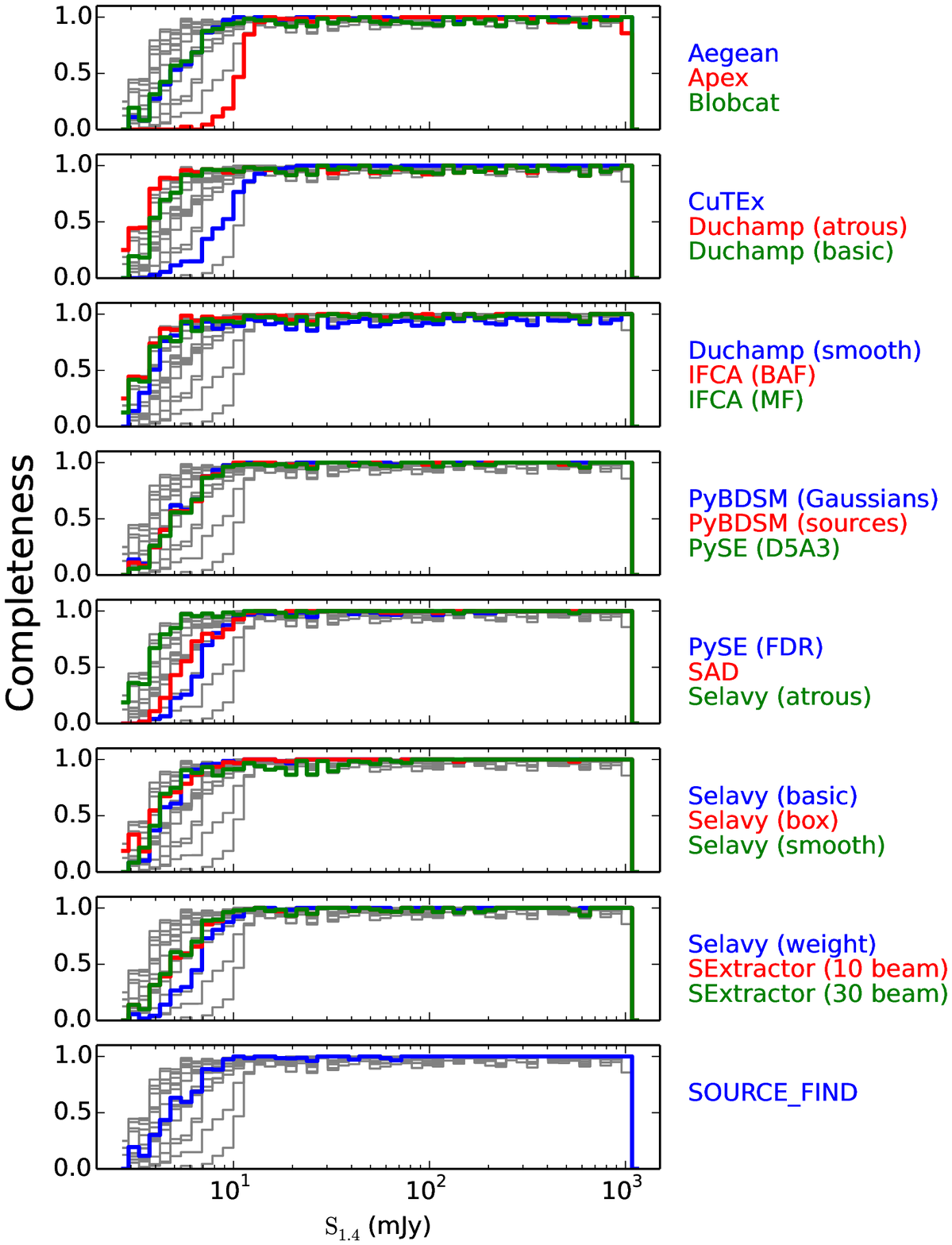}\hspace{-2mm}
\includegraphics[width=9cm]{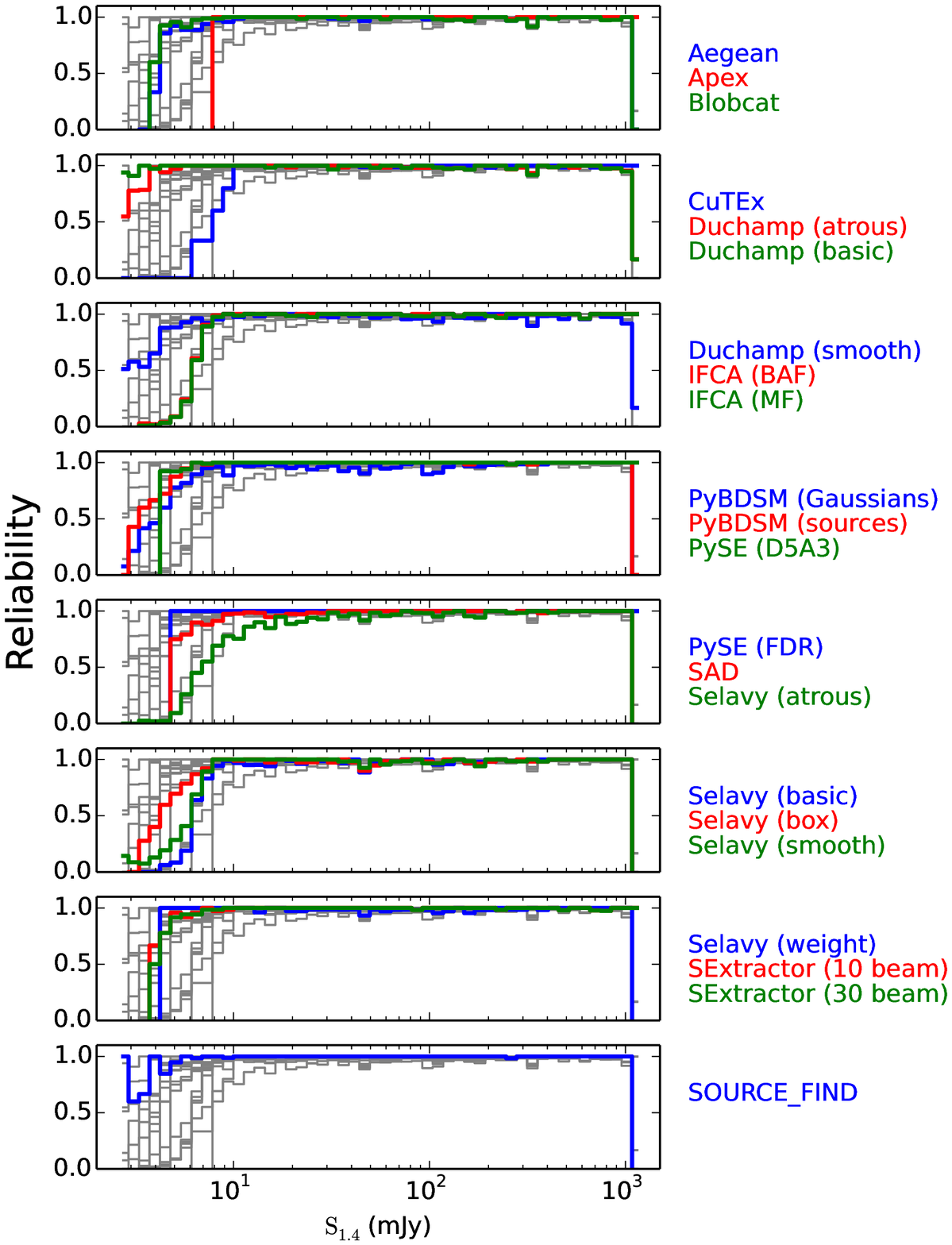}}
\caption{The completeness and reliability fractions (left and right respectively) as a function of
input source flux density (completeness) or measured source flux density (reliability) for each of the tested
finders for Challenge~2. The grey lines
show the distribution for all finders in each panel, to aid comparison for any given finder.}\label{completeness2}
\end{center}
\end{figure*}

\begin{figure*}[ht]
\begin{center}
\centerline{\includegraphics[width=9cm]{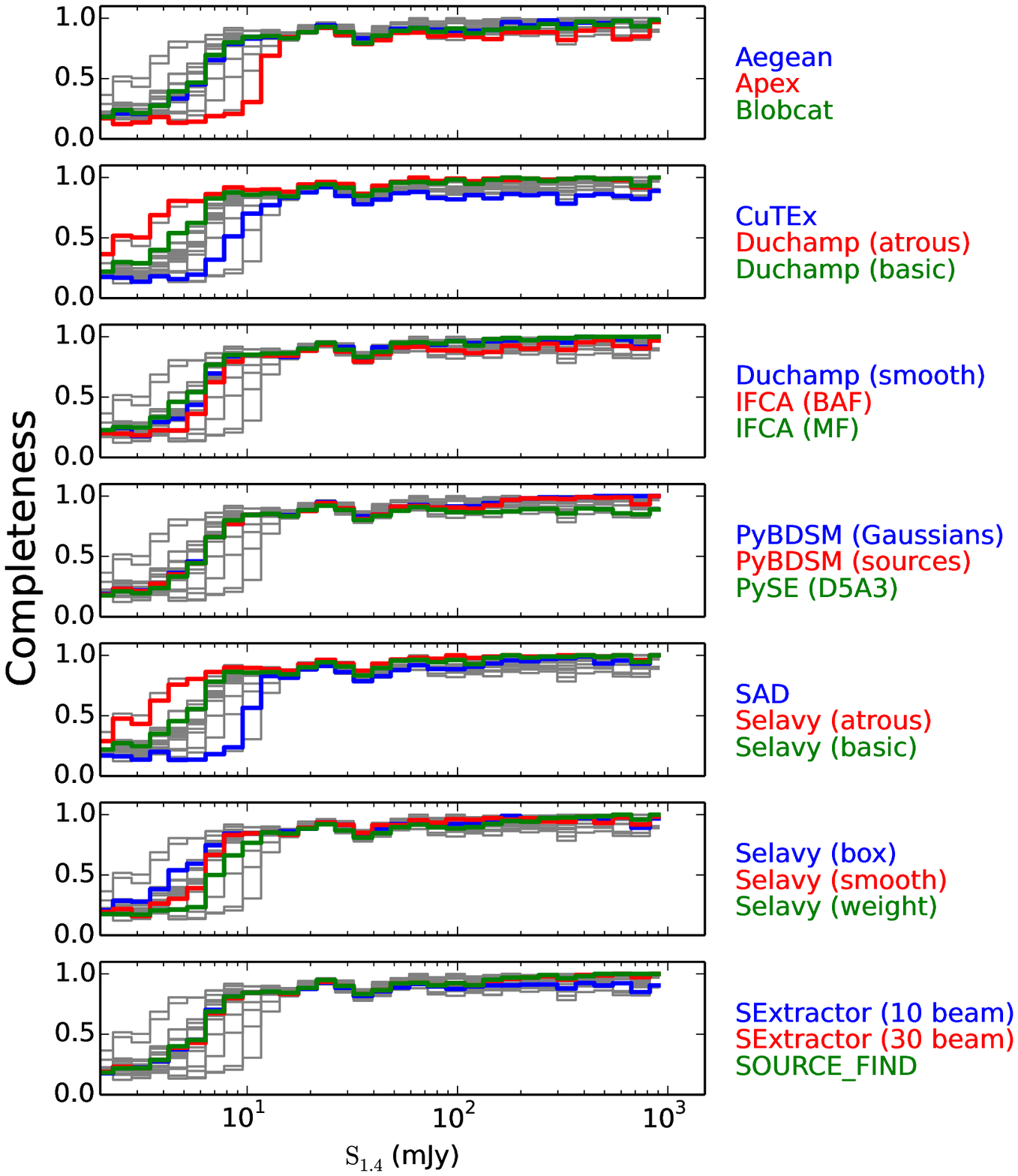}\hspace{-2mm}
\includegraphics[width=9cm]{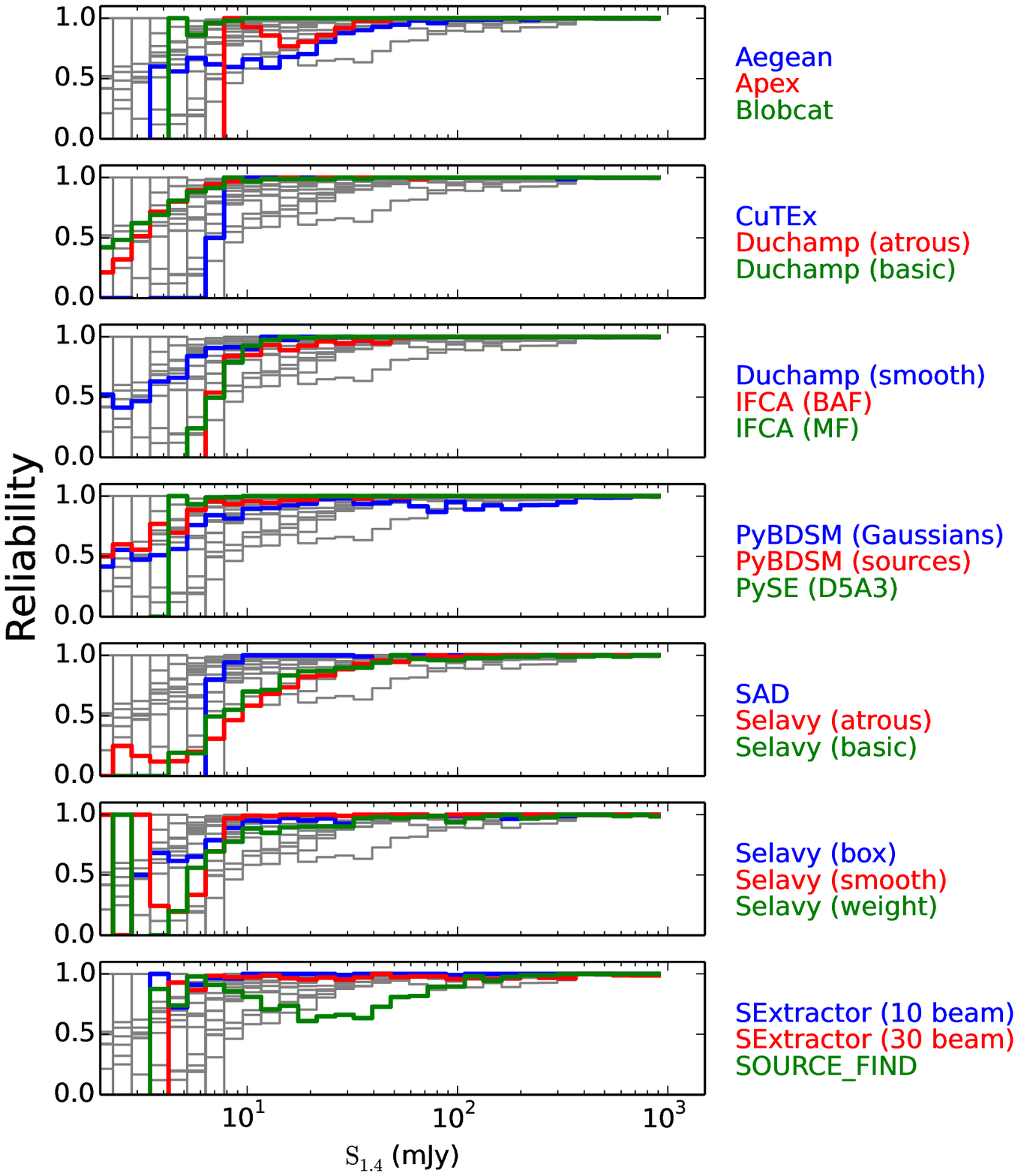}}
\caption{The completeness and reliability fractions (left and right respectively) as a function of
input source flux density (completeness) or measured source flux density (reliability) for each of the tested
finders for Challenge~3. The grey lines
show the distribution for all finders
in each panel, to aid comparison for any given finder. Note that PySE (FDR) was only submitted
for Challenges~1 and 2, and does not appear here.}\label{completeness3}
\end{center}
\end{figure*}

Completeness and reliability are commonly used statistics for a measured set of sources
to assess the performance of the source finder. The completeness is the fraction of real (input) sources
correctly identified by the measurements, and the reliability is the fraction of the measured sources
that are real.

To compare the submitted results for each finder with the input source lists, we first perform
a simple positional cross-match. For Challenges~1 and 2 we define a measured source
to be a match with an input source if it is the closest counterpart with a positional offset
less than $5''$. This offset corresponds to 2.5 pixels or about $0.5\times$FWHM of the resolution
element, so is a suitably small offset to minimise false associations. By construction there are
no input sources within this positional offset of each other, ensuring that any match with a
measured source should be a correct association.
For Challenge~3, given the presence of extended sources, we increased
the offset limit to $30''$, roughly $3\times$FWHM of the resolution element, to account for
greater positional uncertainties in the detected sources by the different finders.
This does lead to the possibility of spurious cross-matches between measured and
input sources. We do not attempt to account for this effect in the current analysis, though,
merely noting that this is a limitation on the accuracy of these metrics for Challenge 3, and
that any systematics are likely to affect the different finders equally. We avoid
using additional criteria such as flux density \citep[e.g.,][]{Wan:14} to refine the cross-matching,
as this has the potential to conflate our analyses of positional and flux density accuracy.
Using this definition, we calculate the completeness and reliability
of the input catalogues for each of the three Challenges. These are shown in
Figures~\ref{completeness1}, \ref{completeness2} and \ref{completeness3}.

We show these measurements as a function of the input source flux density for the completeness
measure and of the measured source flux density for the reliability. Ideally, the
S/N rather than the flux density should be used here, but because of the way the
artificial images have been generated, following a simulated observation and deconvolution process,
the intrinsic S/N is not known {\em a priori}. We measure the root-mean-square (rms) noise level
in the images directly, at several representative locations selected to avoid bright sources. We note
that the unit of flux density in each pixel is mJy/beam, so that changing the pixel scale in the image
changes only the number of pixels/beam, not the flux scaling. We measure $\sigma \approx 9\,$mJy for
Challenge~1 and $\sigma \approx 1\,$mJy for each of Challenge~2 and 3,
although the value fluctuates as a function
of location in the image by up to $\pm2\,$mJy for Challenge~1 and  $\pm0.5\,$mJy
for Challenges~2 and 3. Bearing these details in mind, much of the discussion below refers in general
terms to S/N rather than to flux density, in order to facilitate comparisons between the Challenges.

The completeness and reliability curves give insights into the performance of the various finders.
In broad terms, most finders perform well at high S/N, with declining completeness and reliability below
about $10\,\sigma$. In general we see the expected trade-off between completeness and reliability,
with one being maintained at the expense of the other, but there are clearly variations of
performance between finders and Challenges. It may be desirable in some circumstances to use certain metrics
(such as the S/N at which completeness drops to 50\%, or the integrated reliability above some threshold) to
summarise the information contained in the completeness and reliability distributions. Due to the nature of the current
investigation, though, and in order not to obscure any subtle effects, we have chosen to focus on the
properties of the full distributions.

For all finders, the qualitative performance in Challenge~1 is similar to the performance
in Challenge~2, although quantitatively the completeness and reliability are poorer in Challenge~1
than in Challenge~2. Finders that demonstrate a good performance at low S/N in terms of completeness
while also maintaining high reliability include Aegean, {\sc blobcat}, SExtractor and SOURCE\_FIND. IFCA
(in both modes) has a very high completeness, but at the expense of reliability. CuTEx shows the
lowest completeness as well as reliability at faint levels.

Some finders ({\sc blobcat}, Duchamp, and Selavy in smooth mode) at high S/N show
dips or declining performance in either or both of completeness and reliability, where the results should be
uniformly good. At very high S/N we expect 100\% completeness and reliability from all finders.
Some finders that perform well in terms of completeness still show poorer than expected levels of
reliability. Selavy in most modes falls into this category, as does PyBDSM (Gaussians) for Challenge~2 (but not
Challenge~1, surprisingly).

For those finders that otherwise perform well by these metrics, we can make a few more observations. First
it is clear that the APEX finder used a higher threshold than most finders, approximately a $10\,\sigma$
threshold compared to something closer to $5\,\sigma$ for all others. Is it also apparent that SAD
demonstrates a drop in reliability below about $10\,\sigma$ that declines faster than most of the
other similarly-performing finders, before recovering at the lowest S/N at the expense of completeness.
This is emphasised more in Challenge~1 than in Challenge~2.

The performance of most finders in Challenge~3 is similar to that in other Challenges, except for a reduced
completeness and reliability. This is not surprising as the 20\% of sources that are extended will have a
reduced surface brightness and hence peak flux density compared to their total flux density, so
many of the extended sources are below the threshold for detection for all the finders tested. In addition, the
reliability is likely to be reduced at low to modest S/N as a consequence of the extended emission from these
sources pushing some noise peaks above the detection threshold. This may also arise from the number of
artifacts related to the extended sources that are visible in Figure~\ref{challenge3}.
Most finders still demonstrate a completeness for Challenge~3 of better than around 80\% above reasonable flux density
(or S/N) thresholds (e.g., S/N$\ge 8 - 10$), which is encouraging since this is the fraction of input sources in
Challenge~3 that are
point sources. Despite this, {\sc blobcat}, PyBDSM (sources), PySE (D5A3), SAD and SExtractor maintain very high
reliability in their measured sources for Challenge~3. Other finders, though, including Aegean and SOURCE\_FIND,
as well as Selavy, show very low reliability in this Challenge, even at very high S/N, suggesting that there
may be additional issues contributing to detection of false sources in the presence of extended sources.
We note that these finders are designed for the detection of point sources, but further investigation is
needed to establish why the presence of extended emission affects their performance in this way.
One possibility is that an extended source may be broken into a number of individual point-source components,
due to noise fluctuations appearing as local maxima. These would then appear as false detections since they
do not match up to an input source.

\begin{figure*}[ht]
\begin{center}
\centerline{\includegraphics[width=6cm]{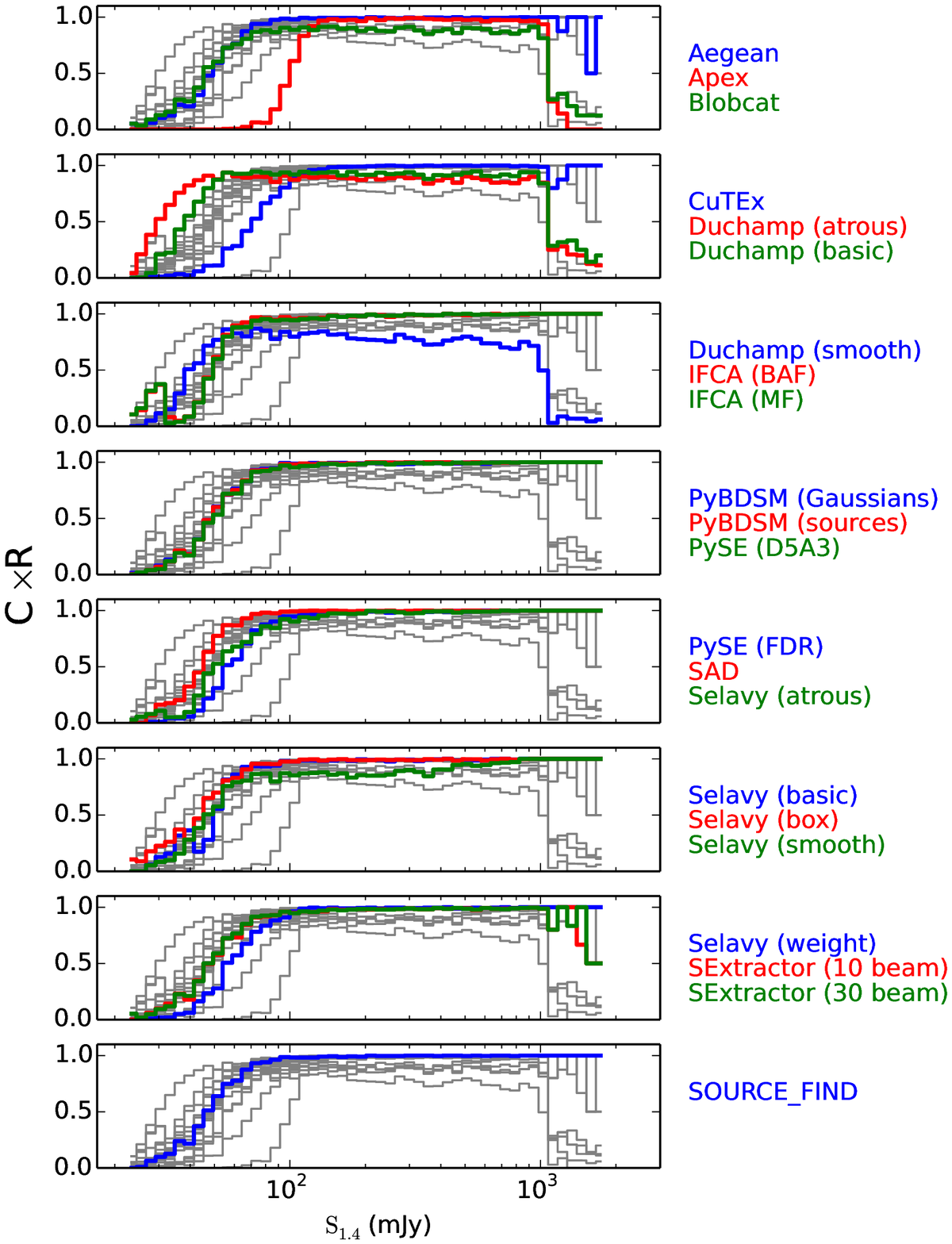}\hspace{-2mm}
\includegraphics[width=6cm]{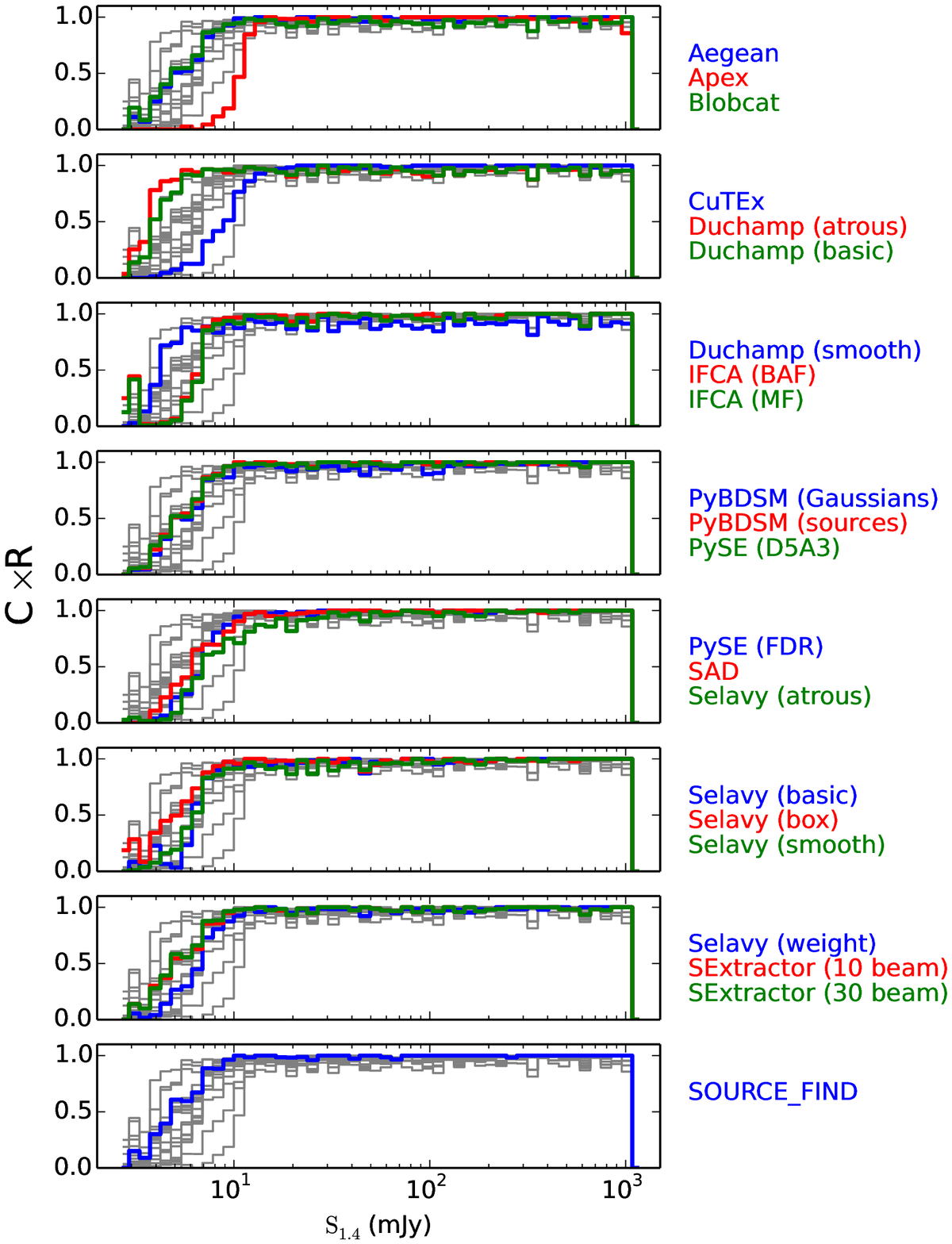}\hspace{-2mm}
\includegraphics[width=6cm]{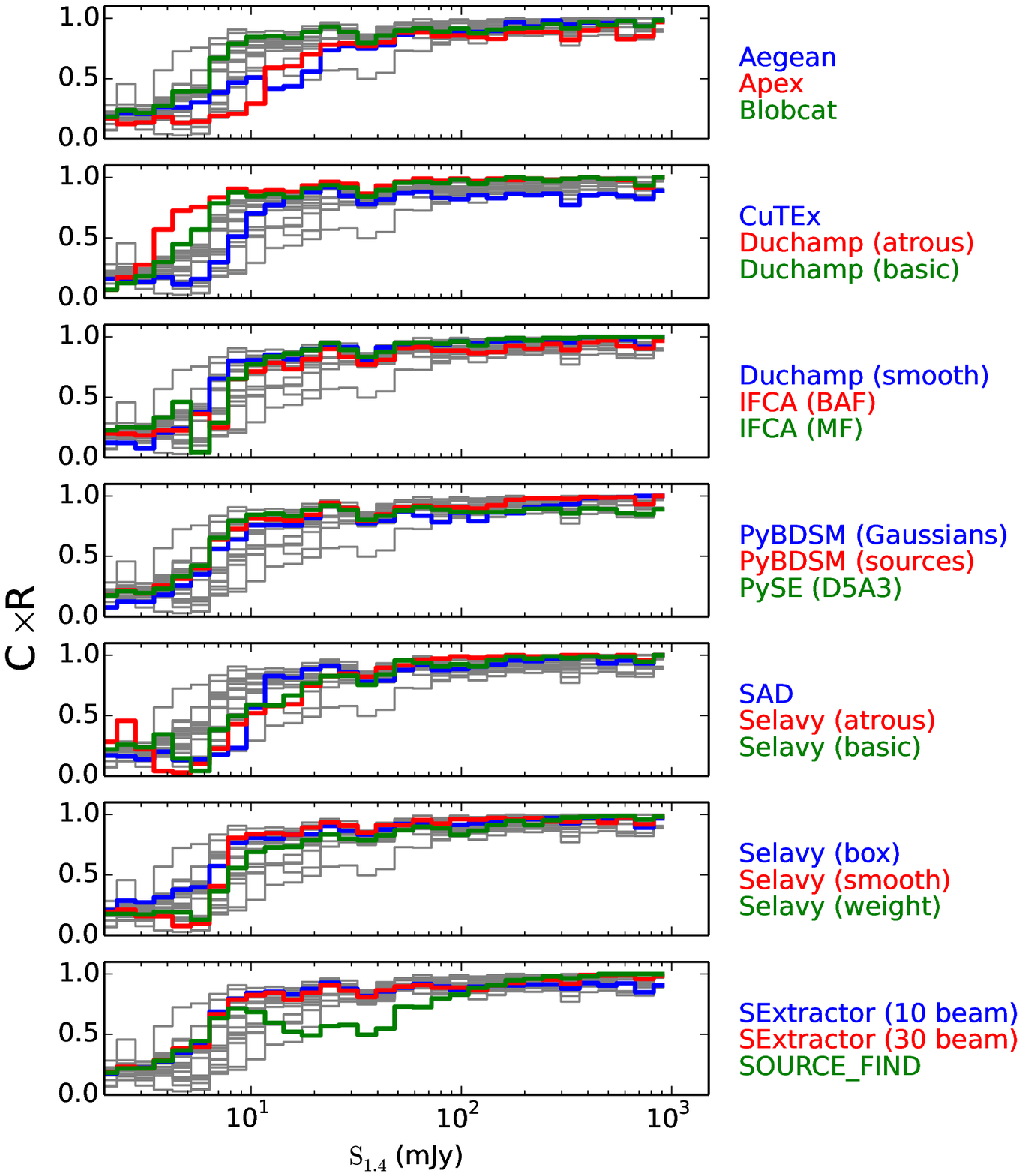}}
\caption{The product of the completeness and reliability as a function of input source flux density for each of the
tested source finders for Challenges~1-3 (left to right). The grey lines show the distribution for all
finders in each panel, to aid comparison for any given finder. Note that PySE (FDR) was only submitted
for Challenges~1 and 2.}\label{CRprod}
\end{center}
\end{figure*}

Since maximising both completeness and reliability is one clear goal of source finding, we illustrate in
Figure~\ref{CRprod} how the product of completeness and reliability for all finders varies as a function of the
input source flux density for each of the three Challenges. The advantage of this metric is that it retains
the dependence on flux density (or S/N), so that the joint performance can be assessed as a function of
source brightness. This may serve to provide a more direct or intuitive comparison
between finders at a glance than the separate relationships from Figures~\ref{completeness1}, \ref{completeness2}
and \ref{completeness3}. It readily highlights finders than perform poorly at high S/N (e.g., {\sc blobcat} and Duchamp
in Challenge~1), and the range of performance at low S/N. It also highlights that most finders follow a quite tight
locus in this product as the flux density drops from about $10\,\sigma$ to $5\,\sigma$ and below, and can be used
to identify those that perform better or worse than this typical level.
Clearly, though, the origin of any shortfall in the product of the two statistics needs to be identified in the earlier Figures.

\begin{figure*}[ht]
\begin{center}
\centerline{\includegraphics[width=6cm, angle=0]{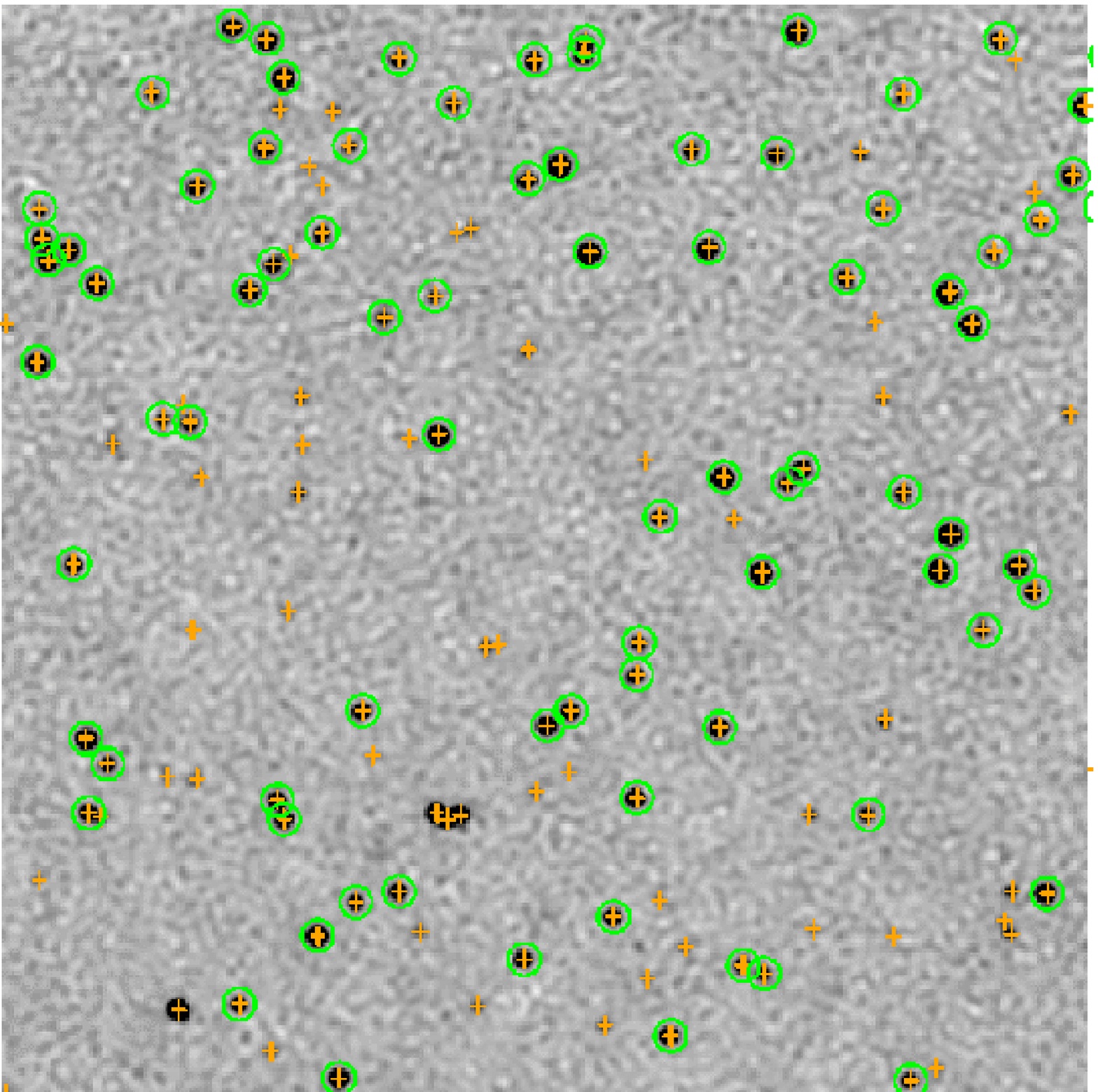}\hspace{5mm}
\includegraphics[width=6cm, angle=0]{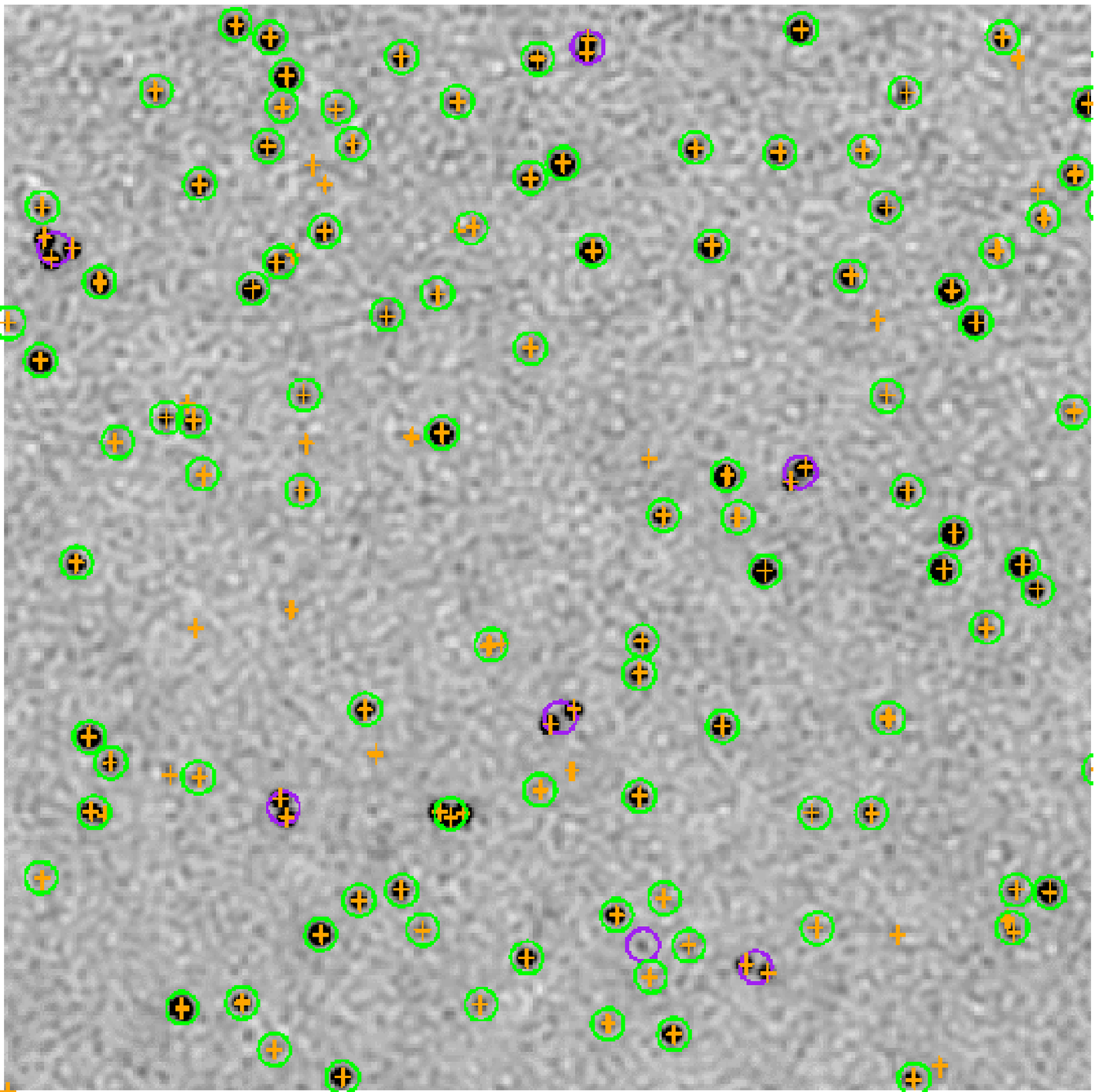}}
\vspace{4mm}
\centerline{\includegraphics[width=6cm, angle=0]{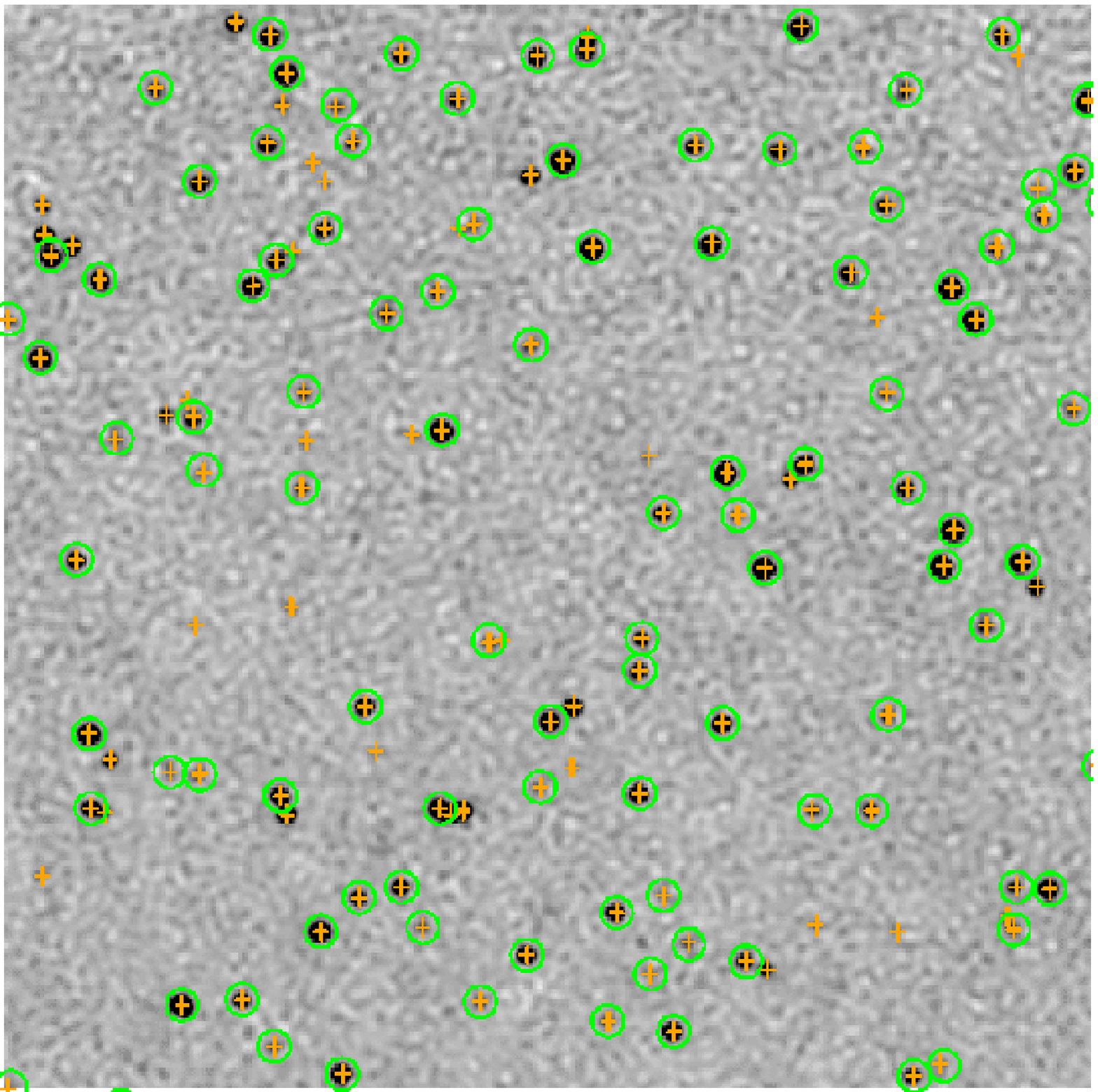}\hspace{5mm}
\includegraphics[width=6cm, angle=0]{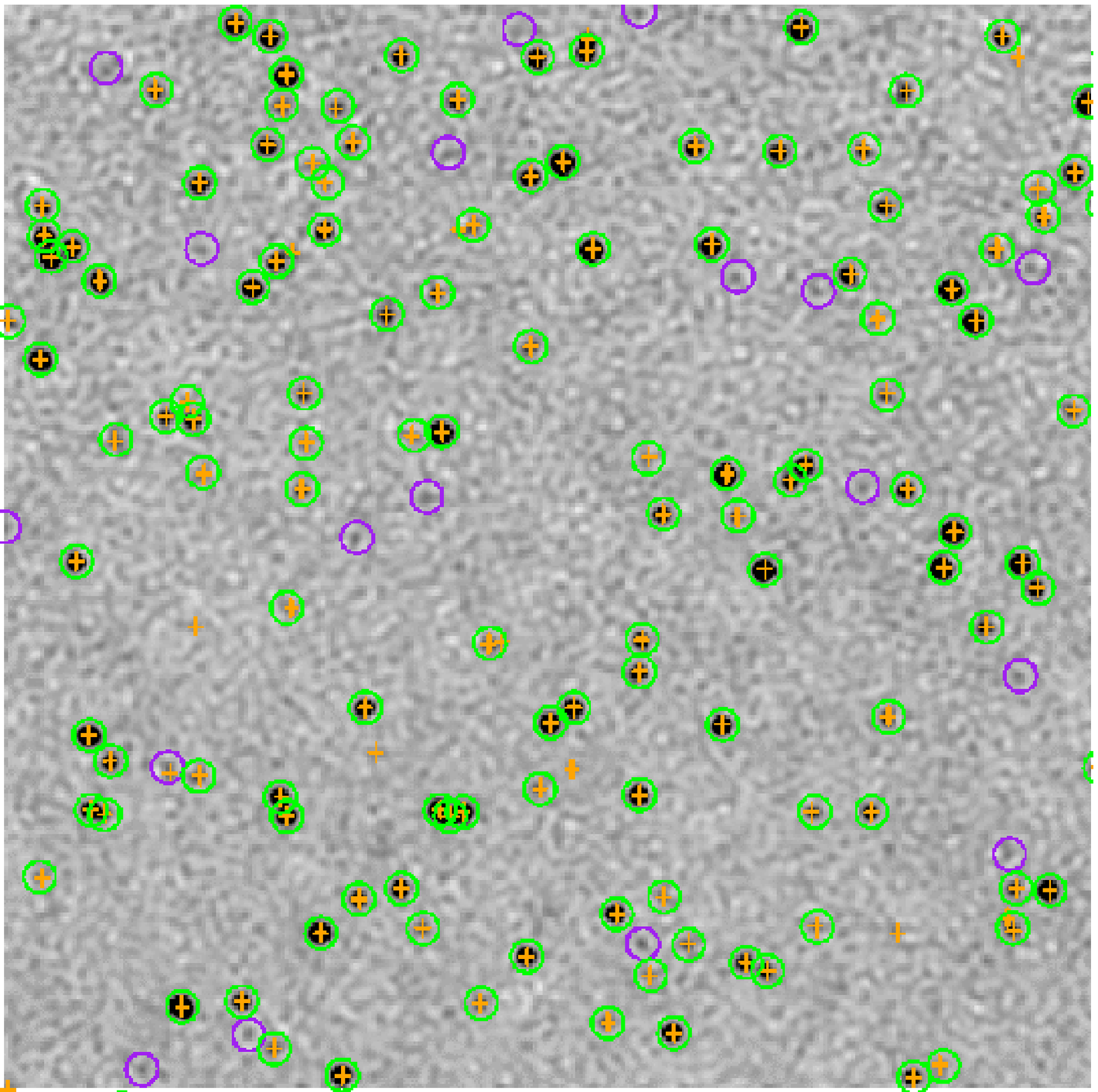}}
\caption{Examples illustrating potential sources of both incompleteness and poor reliability for four of the
tested source finders for Challenge~1. Top left: Apex; Top right: {\sc blobcat}; Bottom left: Selavy (smooth);
Bottom right: Selavy (atrous). Orange crosses identify the location of input artificial sources. Circles are the
sources identified by the various finders, with green indicating a match between a measured source
and an input source, and purple indicating no match. Isolated orange crosses indicate incompleteness,
and purple circles show poor reliability.}\label{examples1}
\end{center}
\end{figure*}

There is an issue related to source blending that affects the degree of completeness
reported for some finders. This is evident in particular for significantly bright sources
which all finders should detect, and is for the most part a limitation in the way the
completeness and reliability is estimated based on near-neighbour cross-matches.
The practical assessment of completeness and reliability is problematic in particular for finders that use
a flood-fill method and do not do further component fitting.
Both {\sc blobcat} and Duchamp merge sources if the threshold is sufficiently low, and then report the merged
object. This is likely the origin of their apparently poor performance at high S/N in Challenge~1, where many bright
sources may be overlapping. If the centroid or flux-weighted position reported for the merged object lies further
from either input source location than the matching radius used in assessing counterparts between the
input and submitted source lists, the detected blend will be excluded.
Note that the higher spatial density of sources
at bright flux density in Challenge~1 makes this more apparent than in Challenge~2 (compare
Figures~\ref{completeness1} and \ref{completeness2}).
While this seems to be the cause of most of the issues,
there are clearly some cases where bright sources are genuinely missed by some finders.
Figure~\ref{examples1} shows that Selavy (smooth)
has a tendency not to find bright sources adjacent to brighter, detected sources. Selavy (\`a trous),
on the other hand, does detect these but at the expense of finding many more spurious sources. This is
discussed further below.

Figure~\ref{examples1} provides an illustration of some of the sources of incompleteness and
reliability. The issue of blended sources being detected but reported as a single object by {\sc blobcat}
can be easily seen in this example. Here 5 adjacent pairs and 1 adjacent triplet are successfully
detected by {\sc blobcat} but reported with positions, from the centroid of the flux distribution,
sufficiently different from the input catalogue that they are not recognised as matches. This is
likely to be the cause of the apparent reduction in completeness for {\sc blobcat} at the higher flux density
levels. We note that {\sc blobcat} provides a flag to indicate when blobs are likely to consist of blended
sources. These flags were not followed up for deblending in the submitted results due to the stated focus
in this Challenge on point-like sources. It is clear, though, that even for point sources the blending issue needs
careful attention. In a production setting, automated follow-up of blended sources will be required to
improve completeness. The effectively higher flux density (or signal-to-noise) threshold
of Apex is also visible in this Figure as the large number of sources not detected.

The two Selavy results shown in Figure~\ref{examples1} also provide insight into the possible failure modes of finders.
The two algorithms shown are the ``smooth" and ``\`a trous" methods. The smooth approach smooths the image with
a spatial kernel before the source-detection process of building up the islands, which are then fitted with 2D Gaussians.
In some cases multiple components will have been blended into a single island, which is clearly only successfully
fitted with one Gaussian. This leads to one form of incompleteness, probably explaining the lower overall completeness
for this mode in Figure\,\ref{completeness1}. The \`a trous approach reconstructs the image
with wavelets, rejecting random noise, and uses the reconstructed image to locate the islands that are subsequently
fitted. This gives an island catalogue with nearby components kept distinct (which are each subsequently fit by a
single Gaussian), but has the effect of identifying more spurious fainter islands (note the larger number of purple circles
compared to the Selavy smooth case), leading to the poorer reliability seen in Figure\,\ref{completeness1}.
This analysis demonstrates the importance of both the detection and fitting steps for a successful finder.

\subsection{Image-based accuracy measures}
\label{iman}

The method of calculating the completeness and reliability based on source identifications
depends critically on the accuracy of the cross-matching. This can be problematic in the event
of source confusion, where distinct sources lie close enough to appear as a single, complex
blob in the image. As an alternative approach for evaluating the completeness we create
images from the submitted catalogues, and compare on a pixel-by-pixel basis with the challenge
images and original model images (smoothed with the same beam as the challenge images).
This provides a way of assessing how well the different finders replicate the distribution of
sky brightness for the image being searched. This approach may favour over-fitting of sources,
but still provides an important complement to the analysis above.

The images are made using the same technique described in Section~\ref{sec-images}.
Where available the measured size of the source components was used, but if only the
deconvolved size was available the size was convolved with the beam according to the
relations in \citet{Wil:70}. This produced Gaussian blobs that were distributed onto the
same pixel grid as the input images to create the ``implied image''. As before, the images are
cropped to remove the regions where the image noise was high. We consider residual images
made in two ways, by subtracting either the Challenge image or the original
input model from this implied image. The following statistics were
calculated from the residual: the rms; the median absolute deviation from
the median \citep[MADFM, which we convert to an equivalent rms by dividing
by 0.6744888;][]{Whi:12}; and the sum of the squares. The latter and the rms
provide an indication of the accuracy including outliers (which will be from sources
either missed or badly fit), while the MADFM gives an indication of where the bulk of the
residuals lie. A finder that fits a lot of sources well, but still has a few poor fits, will tend to have
a lower MADFM value but somewhat higher rms and sum of squares. The results are shown in
Tables~\ref{imAnalysis1}, \ref{imAnalysis2} \& \ref{imAnalysis3}
for Challenges~1, 2 and 3 respectively.

These statistics are related to the properties of the noise in both the Challenge and implied images. To
address this in order to have some benchmark for the measured values, we perform the analysis on each of
the Challenge images themselves by subtracting the Challenge image from the smoothed model.
This gives the measurements that would correspond to optimal performance if a finder recovered
the full input catalogue, given the (false) assumption that the noise is identical between the Challenge image
and each implied image (i.e., some of the difference between finder metrics and the benchmark value will
be attributable to noise). We note that this benchmark is only relevant for the metrics calculated by
subtracting the Challenge image from each implied image.
Because the benchmark is limited by the absence of noise in the smoothed model, we treat
this analysis as a relative comparison between finders rather than as an absolute metric.

{\sc blobcat} does not provide shape information in the form of a major and minor axis with a position angle.
Although the ratio between integrated to peak surface brightness can be used to estimate a characteristic angular size
(geometric mean of major and minor axis), and these values were provided in the submission to the
Data Challenge, they do not allow for unambiguous reconstruction of the flux density distribution and we
have not included these in the present analysis so as to avoid potentially misleading results.

Different finders assume different conventions in the definition of position angle. We strongly recommend
that all source finders adopt the IAU convention on position angle to avoid ambiguity. This states that
position angles are to be measured counter-clockwise on the sky, starting from North and increasing
to the East \citep{IAU:74}. We found that we needed to rotate Aegean's position angles by $90^\circ$ (an early error
of convention in Aegean, corrected in more recent versions), and to reverse the sign of the IFCA position
angles, to best match the input images. In the absence of these modifications cloverleaf patterns, with
pairs of positive and negative residuals at $\sim90^\circ$ from each other, appear at the location of
each source in the residual images. CuTEx position angles were not rotated for this analysis, although
we found that similar cloverleaf patterns existed on significantly extended components, most notable
in Challenge~3. If we adjusted the position angle, though, cloverleaf residuals appeared on the more
compact sources. The non-rotated catalogue performs better than the rotated version, although for completeness
we report both versions in Tables~\ref{imAnalysis1}, \ref{imAnalysis2} and \ref{imAnalysis3}. The CuTEx flux densities
as submitted were systematically high by a factor of two, and subsequently corrected after identifying a trivial post-processing
numerical error (see \S\ref{fluxes} below). The analysis here has accounted for this systematic by dividing the
reported CuTEx flux densities by two.

The finders that generally appear to perform the best in this analysis are Aegean, CuTEx and PyBDSM.
For PyBDSM the Gaussians mode seems to perform better than the Sources mode, across the three Challenges,
although the performance for both is similar apart from Challenge~3 where the Gaussians mode is more appropriate
for estimating the properties of the extended sources. The finders that seem to most poorly reproduce the flux distribution
are SExtractor and IFCA. We discuss this further below in \S\ref{fluxes}.

Selavy performs reasonably well in these tests, with results generally comparable to the better
performing finders. It is worth noting, though, that the different Selavy modes perform differently
in each Challenge. For Challenge~1, Selavy (\`a trous) performs best; for Challenge~2 it is Selavy (smooth);
and for Challenge~3, Selavy (basic) and Selavy (box) both perform well. This can be understood in
terms of the different source distributions and properties in each of the three Challenges. The
high density of bright sources in Challenge~1 seems to be best addressed with the \`a trous mode, the Smooth
mode for the more realistic source distribution of Challenge~2, and the extended sources of Challenge~3
better characterised by the Basic and Box approaches. This leaves an open question over which
mode is better suited to the properties of sources in the real sky, and this will be explored as one of
the outcomes from the current analysis.

\subsection{Positional and flux density accuracy}

In addition to the detection properties, we also want to assess the characterisation accuracy of
the different finders. This section explores their performance in terms of the measured positions
and flux densities. Because not all finders report size information, we chose not to
include measured sizes in the current analysis. Because the restoring beam in the Challenge
images was close to circular, we also chose not to investigate position angle estimates.

\subsubsection{Positions}
\label{positions}

In order to assess positional accuracy we use the relationships defined by \citet{Con:97} that establish the
expected uncertainties from Gaussian fits \citep[see also][]{Hop:03}. These relations give expected
positional errors with variance of
\begin{equation}
\mu^2(x_0)\approx (2\sigma_x)/(\pi \sigma_y)\times(h^2\sigma^2/A^2),
\end{equation}
\begin{equation}
\mu^2(y_0)\approx (2\sigma_y)/(\pi \sigma_x)\times(h^2\sigma^2/A^2),
\end{equation}
where $\sigma$ is
the image rms noise at the location of the source, $h$ is the pixel scale and $A$ is the amplitude of
the source. The parameters $\sigma_x$ and $\sigma_y$ are the Gaussian $\sigma$ values of the source in the $x$ and $y$
directions. Here, $\theta_M$ and $\theta_m$, the full width at half maximum along the major and minor axes, can
be interchanged for $\sigma_x$ and $\sigma_y$, as the $\sqrt{8\ln 2}$
factor cancels. If the source size is $n$ times larger in one dimension than the other,
the positional error in that dimension will be $n$ times larger as well. In our simulations,
the point sources in the images arise from a point spread function that is approximately
circular, and $\theta_M \approx \theta_m$. Correspondingly, the positional rms errors in both
dimensions should be $\mu \approx \sqrt{2/\pi}(h\sigma/A)$. For our simulated images $h=2''$, and
accordingly we would expect the positional errors from a point source finder due to Gaussian noise alone
to be $\mu \approx 0.3''$ for S/N=5, $\mu \approx 0.15''$ for S/N=10, and $\mu \approx 0.05''$ for S/N=30.

\begin{table*}[h]
\caption{Results from image-based analysis, for Challenge~1. We consider residual images made in two ways, subtracting
either the image or the smoothed model from the implied image, and measure the rms derived from the MADFM (in
mJy/beam), and the sum of the squares of the residuals (in (Jy/beam)$^{2}$). We show for comparison, in the line
labelled ``input", the same statistics derived
from subtracting the smoothed model from the challenge image. In each column the three submitted entries with the lowest
values are highlighted in bold, as is the best performance of Selavy for reference.}
\begin{center}
\begin{tabular}{l|rr|rr}
   &\multicolumn{2}{c}{Image} &\multicolumn{2}{c}{Model}\\
ID &MADFM &sumsq & MADFM & sumsq\\ \hline 
Input &9.4570 &2910.7 & --- &---\\ \hline  
APEX &9.5478 &4351.9 & 0.0261 &1481.4\\ 
Aegean & \textbf{9.4242} & \textbf{2969.5} & 0.0258 &\textbf{186.3}\\ 
CuTEx &9.5119 &3147.5 & 0.0270 &312.7\\ 
CuTEx (rotated) &9.5272 &3153.9 & 0.0272 &323.0\\ 
IFCA BAF &9.5641 &11433.0 & 0.0280 &8925.0\\ 
IFCA MF &9.7680 &35221.0 & 0.0306 &33483.0\\ 
PyBDSM gaussians &\textbf{9.3271} &\textbf{2920.6} & 0.0260 &\textbf{162.4}\\ 
PyBDSM sources & \textbf{9.3395} & \textbf{2976.3} & 0.0260 & \textbf{216.1}\\ 
PySE D5A3 &9.4389 &3098.3 & \textbf{0.0252} &263.4\\ 
PySE FDR &9.4546 &3119.2 & \textbf{0.0254} &279.0\\ 
SAD &9.4777 &3020.2 & 0.0257 &224.9\\ 
SExtractor 10 beam &9.6956 &7566.8 & 0.0302 &4890.1\\ 
SExtractor 30 beam &9.6915 &7570.5 & 0.0302 &4897.1\\ 
SOURCE\_FIND &9.5116 &3099.0 & \textbf{0.0254} &286.4\\ 
\hline
Duchamp \`a trous &9.6905 &4609.4 & 0.0300 &1922.3\\ 
Duchamp basic &9.7371 &4234.1 & 0.0297 &1430.7\\ 
Duchamp smooth &9.7143 &6594.3 & 0.0287 &3812.0\\ 
Selavy \`a trous &\textbf{9.2033} &\textbf{2791.1} & 0.0318 &324.2\\ 
Selavy basic & 9.3207 & 2897.5 & 0.0265 &196.0\\ 
Selavy box &9.3330 &2923.3 & 0.0259 &\textbf{165.2}\\ 
Selavy smooth &9.4829 &4251.1 & \textbf{0.0253} &1405.1\\ 
Selavy weight &9.3802 &2994.0 & 0.0259 &211.9\\ 
\end{tabular}
\end{center}
\label{imAnalysis1}
\end{table*}

\begin{table*}[h]
\caption{Results from image-based analysis, for Challenge~2. Columns as for Table.~\ref{imAnalysis1}.}
\begin{center}
\begin{tabular}{l|rr|rr}
   &\multicolumn{2}{c}{Image} &\multicolumn{2}{c}{Model}\\
ID & MADFM &sumsq & MADFM & sumsq\\ \hline 
Input & 1.0391 &35.4 & --- &---\\ \hline  
APEX & 1.0425 &123.4 & \textbf{0.0044} &88.2\\ 
Aegean & 1.0404 &\textbf{38.5} & 0.0045 &\textbf{3.6}\\ 
CuTEx & 1.0419 &\textbf{39.7} &0.0045 &\textbf{4.6}\\ 
CuTEx (rotated) & 1.0438 &47.2 & 0.0045 &11.7\\ 
IFCA BAF & \textbf{1.0381} &278.2 & 0.0050 &245.9\\ 
IFCA MF & 1.0452 &1151.8 & 0.0053 &1129.1\\ 
PyBDSM gaussians & \textbf{1.0380} &\textbf{35.8} & 0.0045 &\textbf{0.9}\\ 
PyBDSM sources & \textbf{1.0384} &43.0 & 0.0045 &8.1\\ 
PySE D5A3 & 1.0403 &59.6 & \textbf{0.0044} &24.5\\ 
PySE FDR & 1.0407 &45.9 & \textbf{0.0044} &10.8\\ 
SAD & 1.0420 &43.3 & 0.0045 &8.3\\ 
SExtractor 10 beam & 1.0446 &160.1 & 0.0045 &125.5\\ 
SExtractor 30 beam & 1.0445 &160.0 & 0.0045 &125.5\\ 
SOURCE\_FIND & 1.0423 &46.0 & \textbf{0.0044} &11.1\\ 
\hline
Duchamp \`a trous & 1.0478 &193.9 & 0.0046 &159.0\\ 
Duchamp basic & 1.0489 &186.2 & 0.0045 &150.6\\ 
Duchamp smooth & 1.0474 &239.3 & 0.0045 &203.7\\ 
Selavy \`a trous & \textbf{1.0312} &41.1 & 0.0049 &8.8\\ 
Selavy basic & 1.0369 &\textbf{38.5} & 0.0046 &4.5\\ 
Selavy box & 1.0381 &42.4 & \textbf{0.0045} &7.6\\ 
Selavy smooth & 1.0388 &\textbf{38.5} & \textbf{0.0045} &\textbf{3.9}\\ 
Selavy weight & 1.0393 &42.5 & \textbf{0.0045} &7.6\\ 
\end{tabular}
\end{center}
\label{imAnalysis2}
\end{table*}

\begin{table*}[h]
\caption{Results from image-based analysis, for Challenge~3. Columns as for Table.~\ref{imAnalysis1}.}
\begin{center}
\begin{tabular}{l|rr|rr}
   &\multicolumn{2}{c}{Image} &\multicolumn{2}{c}{Model}\\
ID & MADFM &sumsq & MADFM & sumsq \\ \hline 
Input & 1.1649 &44.4 & --- &---\\ \hline  
APEX & 1.1918 &130.1 & \textbf{0.0063} &87.1\\ 
Aegean & \textbf{1.1870} &107.1 & \textbf{0.0063} &63.7\\ 
CuTEx & 1.1910 &\textbf{67.8} & 0.0064 &\textbf{25.0}\\ 
CuTEx (rotated) & 1.1924 &\textbf{69.0} & 0.0064 &\textbf{25.8}\\ 
IFCA BAF & 1.1918 &248.6 & \textbf{0.0063} &205.7\\ 
IFCA MF & 1.1951 &2428.8 & 0.0064 &2398.4\\ 
PyBDSM gaussians & \textbf{1.1693} &\textbf{45.3} & 0.0080 &\textbf{2.0}\\ 
PyBDSM sources & \textbf{1.1792} &97.4 & \textbf{0.0062} &53.5\\ 
PySE D5A3 & 1.1901 &86.1 & \textbf{0.0063} &43.1\\ 
SAD & 1.1920 &123.7 & \textbf{0.0063} &79.9\\ 
SExtractor 10 beam & 1.1914 &1293.1 & 0.0065 &1249.3\\ 
SExtractor 30 beam & 1.1890 &1359.3 & 0.0064 &1315.2\\ 
SOURCE\_FIND & 1.1904 &210.5 & \textbf{0.0063} &168.1\\ 
\hline
Duchamp \`a trous & 1.1882 &221.6 & 0.0066 &179.2\\ 
Duchamp basic & 1.1934 &191.7 & 0.0065 &148.2\\ 
Duchamp smooth & 1.1870 &259.6 & 0.0065 &215.9\\ 
Selavy \`a trous & 1.2019 &2738.9 & 0.0066 &2691.4\\ 
Selavy basic & \textbf{1.1800} &\textbf{50.3} &  0.0066 & 8.3\\ 
Selavy box & 1.1819 & 51.4 & \textbf{0.0062} &\textbf{7.9}\\ 
Selavy smooth & 1.1869 &60.2 & 0.0063 &17.7\\ 
Selavy weight & 1.1868 &52.6 & 0.0065 &9.5\\ 
\end{tabular}
\end{center}
\label{imAnalysis3}
\end{table*}

\begin{table*}[h]
\caption{Positional accuracy statistics in arcsec. For a $5\sigma$ detection limit, the minimum rms
error expected is $\mu\approx0.3''$. For $10\sigma$, similar to the threshold for APEX, it is $\mu\approx0.15''$.}
\begin{center}
\begin{tabular}{@{}lcccccccc@{}}
\hline\hline
Source finder & \multicolumn{4}{c}{Challenge~1} & \multicolumn{4}{c}{Challenge~2} \\
 & $\overline{\delta {\rm RA}}$ & $\overline{\delta {\rm Dec}}$ & $\mu_{\rm RA}$ & $\mu_{\rm Dec}$ &
 $\overline{\delta {\rm RA}}$ & $\overline{\delta {\rm Dec}}$ & $\mu_{\rm RA}$ & $\mu_{\rm Dec}$ \\
\hline
Aegean & 0.006 & 0.0005 & 0.53 & 0.52 & $-$0.007 & 0.0005 & 0.68 & 0.65 \\
APEX & $-$0.01 & 0.005 & 0.33 & 0.35 & $-$0.014 & 0.009 & 0.31 & 0.31 \\
{\sc blobcat} & $-$0.008 & $-$0.0003 & 0.70 & 0.71 & $-$0.0003 & $-$0.010 & 0.50 & 0.54 \\
CuTEx & 0.000 & $-$0.0008 & 0.42 & 0.44 & 0.002 & $-$0.005 & 0.35 & 0.39 \\
IFCA BAF & $-$0.0005 & 0.006 & 0.73 & 0.73 &$-$0.006 & 0.012 & 0.63 & 0.63 \\
IFCA MF & $-$0.001 & 0.004 & 0.90 & 0.88 & $-$0.027 & 0.035 & 0.88 & 0.83 \\
PyBDSM Gaussian & 0.005 & 0.0008 & 0.53 & 0.53 & $-$0.002 & 0.004 & 0.51 & 0.52 \\
PyBDSM Source & 0.005 & $-$0.002 & 0.52 & 0.51 &$-$0.003 & $-$0.0009 & 0.45 & 0.44  \\
PySE D5A3 & 0.005 & $-$0.0001 & 0.51 & 0.50 & 0.007 & $-$0.004 & 0.44 & 0.43 \\
PySE FDR & 0.004 & $-$0.002 & 0.39 & 0.39 & 0.005 & $-$0.004 & 0.39 & 0.39 \\
SAD & 0.002 & 0.005 & 0.63 & 0.62 &  0.005 & $-$0.002 & 0.43 & 0.49 \\
SExtractor 10 beam & 1.07 & 0.10 & 2.26 & 2.55 &  0.0007 & $-$0.009 & 0.47 & 0.50 \\
SExtractor 30 beam & 0.0008 & $-$0.0009 & 0.60 & 0.60 & 0.003 & $-$0.009 & 0.48 & 0.50 \\
SOURCE\_FIND & 0.006 & 0.005 & 0.56 & 0.54 & $-$0.0009 & 0.008 & 0.74 & 0.69 \\
\hline
Duchamp \`a trous & $-$0.007 & $-$0.466 & 0.87 & 0.96 & 0.012 & $-$0.487 & 0.72 & 0.87  \\
Duchamp basic & $-$0.007 & $-$0.47 & 0.77 & 0.90 & 0.011 & $-$0.509 & 0.68 & 0.82  \\
Duchamp smooth & $-$0.015 & $-$0.469 & 0.85 & 0.97 & $-$0.005 & $-$0.466 & 0.71 & 0.88  \\
Selavy \`a trous & $-$0.0005 & 0.006 & 0.70 & 0.66 & 0.008 & 0.004 & 0.64 & 0.61 \\
Selavy basic & $-$0.001 & 0.007 & 0.61 & 0.61 &0.010 & $-$0.013 & 0.56 & 0.56 \\
Selavy box & 0.003 & 0.008 & 0.59 & 0.59 & 0.017 & $-$0.008 & 0.58 & 0.56 \\
Selavy smooth & 0.002 & 0.010 & 0.63 & 0.62 & 0.021 & 0.001 & 0.62 & 0.62 \\
Selavy weight & 0.0002 & 0.001 & 0.47 & 0.47 & 0.006 & $-$0.004 & 0.43 & 0.44 \\
\hline\hline
\end{tabular}
\end{center}
\label{posstats}
\end{table*}

The positional accuracies of the finders are presented in Table~\ref{posstats}. We only show the
results for Challenges~1 and 2, since the inclusion of extended sources used in Challenge~3
may result in some fraction of the measured positional offsets arising from real source structure
rather than the intrinsic finder accuracy, making
these simple statistics harder to interpret. Table~\ref{posstats} gives the mean and the rms
in the RA and Dec offsets between the input and measured source positions for each finder. All measured sources
that are in common with input sources are used in calculating these statistics, for each finder. This means that
finders with a higher effective S/N threshold (APEX) should expect to show better rms offsets than the others, since
most sources will have flux densities close to the threshold, and this is indeed the case. For most finders,
with a threshold around S/N$\approx 5$, the best rms positional
accuracy expected would be around $0.3''$. For APEX, with a threshold around S/N$\approx 10$, the best rms
expected should be around $0.15''$. The rms positional accuracies range from a factor of $1.3-2$ larger than
expected from Gaussian noise alone, with CuTEx, PySE and PyBDSM performing the best. SAD performs as well
as these finders in Challenge 2, but not quite as well in Challenge 1.

Almost all of the finders perform well in terms of absolute positional accuracy,
even with the high source density of Challenge~1, with mean positional offsets typically better than 10 milliarcseconds,
or $0.5\%$ of a pixel. A notable exception is SExtractor (10 beam)
in Challenge~1, which has a measurable systematic error in the source positions, and a significantly elevated
rms in the positional accuracy. This is not present for SExtractor (30 beam) or for SExtractor in either mode
for Challenge~2, suggesting that it is a consequence of the high source density present in Challenge~1 and
insufficient background smoothing performed in the 10 beam mode.

For the two finders that we cannot assess blindly, Duchamp shows a systematic positional offset
in Dec, and typically has poorer rms positional errors than most other finders. Selavy generally performs
well, and overall has good mean positions, but has poorer positional accuracy than the best of
the tested finders. The Selavy mode that performs best in terms of rms positional error is Selavy (weight), which is
the mode that performs worst in terms of completeness. This suggests that it may be a lack of low S/N sources that
is causing the estimate of the positional error to appear better. A clear outcome of this test is that Selavy can be
improved by looking at the approach taken by CuTEx, PySE and PyBDSM in estimating source positions.

\subsubsection{Flux densities}
\label{fluxes}
\begin{figure*}[ht]
\begin{center}
\centerline{\includegraphics[width=16cm]{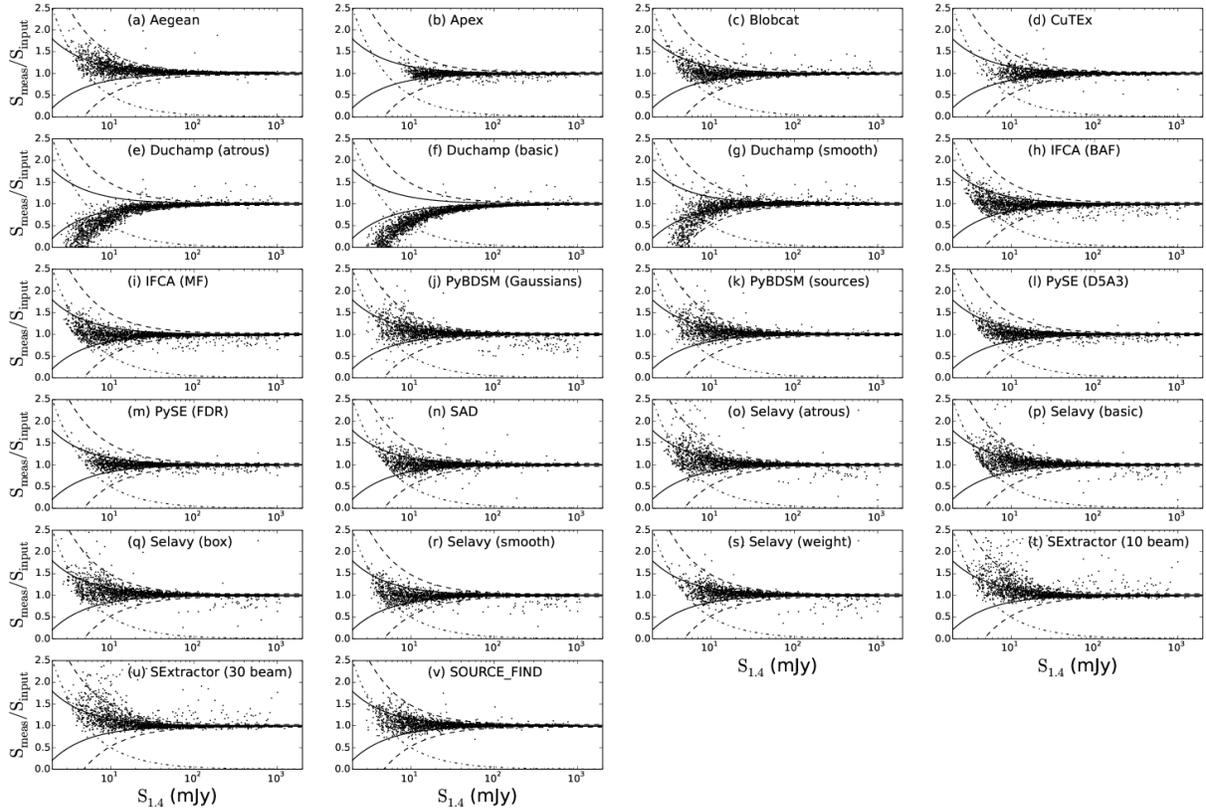}}
\caption{The ratio of the measured to the input flux density, as a function of the input flux density, for
Challenge~2. The solid and dashed lines are the expected $1\sigma$ and $3\sigma$ errors from the
rms noise in the image. The dot-dashed line indicates the expected flux ratio
from a nominal $5\,\sigma$ threshold, obtained by setting ${\rm S_{meas}}=5\,\sigma$ for all values
of ${\rm S_{input}}$.}\label{flxcmp1}
\end{center}
\end{figure*}

The flux density of each component was compared with the input flux density, and
Figure~\ref{flxcmp1} shows the ratio of measured to input flux density as a function of input flux density,
for Challenge~2. Since the sources are point sources, the integrated flux density should be identical
to the peak flux density, but for clarity we use reported peak flux densities from the submissions.
We focus on Challenge~2 as it includes a distribution of input source flux densities
most similar to the real sky. The results from Challenge~1 are similar. We do not consider
Challenge~3 here, again because of the bias introduced by the inclusion of extended sources with low
peak flux densities and high integrated flux densities. Figure~\ref{flxcmp1} indicates with solid and
dashed lines the expected $1\,\sigma$ and $3\,\sigma$ flux density errors respectively, where $\sigma$ here
corresponds to the rms noise level. The dot-dashed line in each panel shows the flux ratio value corresponding
to a $5\,\sigma$ detection threshold. In other words, given the input flux density on the abscissa, the dot-dashed
line shows the ratio that would be obtained if the measured flux density were to correspond to $5\,\sigma$. Values
below (to the left) of this line are only possible if the finder reports measured flux densities for any source
below $5\,\sigma$. This aids in comparison between the depths probed by the different finders.

The need for accurate source fitting is highlighted by the Duchamp results. Duchamp only reports
the flux density contained in pixels above the detection threshold, and so misses a progressively larger
fraction of the flux density as the sources get fainter. For this reason we do not consider Duchamp further
in this discussion of flux density estimation. With the exception of Duchamp, all the finders implement
some form of fitting to account for the total flux density of sources.
They generally show similar behaviours, with reported flux densities largely consistent within the expected
range of uncertainty. The CuTEx flux densities submitted were a factor of two too high, arising from a trivial
numerical error in converting between units of Jy/beam and Jy in the post-processing of the CuTEx output.
This has been subsequently corrected, and
for the remainder of this analysis we consider the CuTEx flux densities after taking this correction into account.

APEX, {\sc blobcat}, IFCA, PySE, and SAD all show flux density errors constrained to within $1\sigma$ for most of the range
of fluxes, and IFCA in particular probes below the nominal $5\,\sigma$ threshold while maintaining an accurate flux
density estimate. Selavy (smooth) performs similarly well. All finders show some fraction of outliers even at
high S/N (S/N$>10$) with flux densities differing as much as $20-50\%$ from the input value. The
fraction of measurements that lie outside the $\pm3\,\sigma$ range is typically a few percent, ranging from $1.8\%$ for
{\sc blobcat} and SOURCE\_FIND, to $4.9\%$ for PyBDSM (Gaussians), and $6.5\%$ for CuTEx after accounting for the factor
of two systematic. SExtractor is notably worse, though, with more than $10\%$ outliers
in both modes. This is likely to result from assumptions about source fitting in optical images, for which SExtractor
was designed, that are less appropriate for radio images. Selavy spans the range of the better performing finders,
with $2.1\%$ outliers for Selavy (smooth) to $4.4\%$ for Selavy (\`a trous).

Catastrophic outliers, with flux densities wrong by $20\%$ or more at high S/N, are more of a concern, especially when
anticipating surveys of many millions of sources. It is possible that some (or even most) of these are related to overlapping
or blended sources, where the reported flux densities are either combined from overlapping sources, or erroneously
assigning flux to the wrong component. Whatever the origin, for most finders the fraction of sources brighter than
30\,mJy (input flux density) with measured flux densities discrepant by $20\%$ or more is $0.2-1\%$.
SExtractor is again a poor performer here, with more than $2\%$ such catastrophic outliers. IFCA ($1.1\%$ in both modes)
and PyBDSM (Gaussians) ($1.9\%$) are notable for also having a larger fraction of such outliers. Aegean, APEX, PyBDSM
(sources), PySE, and SOURCE\_FIND perform the best here, all with around $0.2\%$. Selavy falls in the middle range on
this criterion, with just below $1\%$ catastrophic outliers in all modes.

Aegean, SExtractor, PyBDSM, and to a lesser degree PySE, show a slight tendency to more systematically over-estimate
rather than under-estimate the flux density as the sources became fainter. This is visible in Figure~\ref{flxcmp1} as a
prevalence of $S_{\rm meas}/S_{\rm input}$ values in the $+1\,\sigma$ to $+3\,\sigma$ range and a lack between
$-1\,\sigma$ to $-3\,\sigma$. Another way of saying it is that these finders are, on average, overestimating
the flux densities for these sources, compared to others that do not show this effect. Comparing Aegean and {\sc blobcat},
for example, both have almost identical completeness at these flux densities, implying that the same sources (largely)
are being measured, but while {\sc blobcat} measures
flux densities with the expected symmetrical distribution of $S_{\rm meas}/S_{\rm input}$, Aegean shows an excess
to higher ratios and a deficit at lower. This behaviour is also present for Selavy
in all modes, with the possible exception of Selavy (smooth).

This systematic effect is unlikely to be related to noise bias, where positive noise fluctuations allow a faint source to be
more easily detected, while negative noise fluctuations can lead to sources falling below the detection threshold. That
effect would manifest as a systematic shift above (or below) the dot-dashed threshold locus, not as a deficit of
sources in the $-1\,\sigma$ to $-3\,\sigma$ regime. It is also not related to any imaging biases, such as clean bias
(which typically reduces measured flux densities in any case), because it is not seen in all finders.
It is most likely a consequence of the approach used to perform the Gaussian fitting. At low S/N for point sources there
can be more fit parameters than the data can constrain. The result is that a degeneracy between fit parameters arises,
and it becomes systematically more likely that a nearby noise peak will be mistaken for part of the same source.
So the fit axes are larger and, as a result, the integrated surface brightness also goes up \citep[see Fig.~6 of][]{Hal:12}.

Flux density estimation appears to be more complex, even for simple point sources, than might naively be expected.
While the problem may be mitigated by only fitting point source parameters if the sources are known to be
point-like, in practice this is rarely, if ever, known in advance.
Selavy does not perform especially poorly compared to the other finders tested here, but its performance in all of the aspects
explored above can be improved. None of the tested finders does well in all areas, so specific elements from different finders
will need to be explored in order to identify how best to implement improvements to Selavy and the ASKAPsoft source finder.

\section{DISCUSSION}
\label{disc}
\subsection{Review and comparison of finders}

The purpose of this section is not to identify the ``best" finder in some absolute sense, but rather to summarise the
key outcomes from the above analyses, contrasting the performance of different finders where appropriate, and
highlighting the areas of strength.
Each of the tested finders have strengths and limitations, but none obviously perform best in all of the elements
explored above. Many perform well, while still having individual drawbacks or limitations. Overall, strong performers
include Aegean, APEX, {\sc blobcat}, IFCA, PyBDSM (sources), PySE, and SOURCE\_FIND. A general characteristic
in the completeness and reliability statistics seems to be that finders can maintain high reliability to low S/N
only at the expense of completeness. The most accurate finders follow a similar locus in completeness and reliability
below around $10\,\sigma$ as illustrated in Figure~\ref{CRprod}.

Aegean, {\sc blobcat}, IFCA, PyBDSM (sources) and SOURCE\_FIND all perform similarly well in terms of completeness,
reliability, positional accuracy and flux density estimation. SAD performs well with completeness, positional accuracy
and flux density estimation, but begins to drop away in reliability below about $10\,\sigma$ faster than most other finders.
Aegean has a marginally higher fraction of flux density outliers than
the others, and suffers from the subtle systematic at low S/N to overestimate flux densities. Aegean and SOURCE\_FIND
perform slightly better in terms of reliability at low S/N, but PyBDSM (sources) performs marginally better in terms of
positional accuracy. IFCA in both modes performs similarly in all the elements explored. It shows the highest levels
of completeness among the tested finders at low S/N but this comes at the expense of reduced reliability at these
flux densities. It is also accurate in its position and flux density estimation.

APEX as presented here uses a higher threshold ($\approx 10\,\sigma$) for detection than the other finders. Because of
this, its positional accuracy (Table~\ref{posstats}) is about a factor of two better than nominally expected,
similar in performance to Aegean and SOURCE\_FIND. It also performs similarly well in terms of flux density estimation,
completeness and reliability, to the limits it probes.

PyBDSM performs very differently between the two tested modes. PyBDSM (sources) performs best overall, with good
completeness, reliability, position and flux density estimation. PyBDSM (Gaussians) is poor in terms of reliability for
both Challenges~2 and 3, although it performed well in Challenge~1. Both modes give good positional accuracy, but
PyBDSM (Gaussians) has a relatively large fraction of outliers and catastrophic outliers, in the flux density estimation.
This is likely to be an artifact of our positional cross-matching approach selecting only the nearest submitted source.
PyBDSM may fit a single source by many Gaussians, so if only the closest one is identified as the counterpart to
the input source a lot of the flux density may be artificially missed. The values shown in Tables~\ref{imAnalysis1} and
\ref{imAnalysis2} from the image-based analysis support this conclusion, especially for Challenge~3, suggesting that
PyBDSM is one of the better performers in terms of reproducing the flux distribution in the image. The MADFM and sum
of squares statistics, which are sensitive to outliers, indicate a good performance here.

PySE (D5A3) and PySE (FDR) both provide good positional and flux density estimation, but PySE (FDR) gives marginally
better positions, and is more accurate in flux density estimation with fewer outliers and catastrophic outliers, although
PySE (D5A3) probes to slightly fainter flux densities. PySE (D5A3) performs somewhat better than PySE (FDR) in terms of
completeness, but both are similar in terms of reliability.

CuTEx performs well in terms of positional accuracy and flux density estimation but less well in
completeness and reliability at the low S/N end compared to most of the other finders. We note that
CuTEx was not originally designed to work on radio images but on far-infrared and sub-millimetre images
from space-based facilities, where there is little if any filtering for large scale emission.

Those that perform particularly poorly are SExtractor and Duchamp. SExtractor gives a completeness and reliability that
compare well to most other finders at low S/N, but with flux density estimation that is poorly suited to the
characteristics of radio images. SExtractor was not designed with radio images in mind, and indeed is optimised well
for the Poisson noise characteristics of optical images. It is designed to measure aperture and isophotal magnitudes in a range
of ways that are appropriate for images at optical wavelengths, but understandably these approaches perform more poorly
in the case of radio images when compared to other tools that are designed specifically for that case. 
Duchamp was designed for identifying, but not fitting, sources in neutral hydrogen data cubes rather than continuum
images, and was not expected to perform well in these tests. As expected, it performs poorly in completeness
and reliability, as well as positional and flux density estimation, for well-understood reasons. It has been included in
the current analysis for completeness.

Regarding the performance of Selavy, in the numerous modes tested, we have identified a number of areas for improvement.
Selavy (smooth) performs best in terms of flux density estimation, but is very poor in terms of completeness and reliability.
Selavy (\`a trous) performs better in terms of completeness, but at the expense of very poor reliability and poorer flux
density estimation. The other modes of Selavy are intermediate between these extremes.

\subsection{Common limitations}

Inevitably, all source finders decline in completeness and reliability toward low S/N. It is therefore crucial to quantify
the performance of the finder used for large surveys, in order to associate a well-defined probability of false-detection
with any detected source, and to establish the fraction of sources overlooked at any given S/N level. Future tests like
these will ensure that the ASKAPsoft pipeline is well-calibrated in its behaviour in order to accurately quantify
completeness and reliability.

Positional accuracy of most finders is precise and consistent with the expected uncertainties from Gaussian fits.
However, no finders tested in this Data Challenge perform well in flux density estimation.
As many as $1\%$ of sources at high S/N may have catastrophically poor flux density estimates. These may in part be
associated with blended sources since finders such as Aegean, that do well at deblending, and {\sc blobcat}, that
merge blended sources, show better performance here. Even Aegean and {\sc blobcat} still have $0.2\%$ and $0.4\%$
catastrophic outliers at high S/N, respectively (although note that {\sc blobcat} flags potentially blended sources, see \S\,\ref{CandR}).
For the anticipated catalogues of tens of millions of sources, this will still be a
substantial number of sources. Exploring the origins of and rectifying these catastrophic errors will be an important area of
refinement necessary for the ASKAPsoft source finder, to ensure the high fidelity of the ASKAPsoft pipeline.

\subsection{Updates since the Challenge}

The Data Challenge was completed by the participating teams in early 2013. Since that time many of the source finders
tested in this analysis have had continued development and improved performance. In order to retain the integrity of
the Challenge and to ensure the analysis did not become open-ended, we have chosen to present the results as they
are from the source finders as originally submitted. In order not to provide a misleading representation of the current
state of finders that have improved since that time, we give a brief summary here of some of those developments, and
how they may address any of the limitations identified above.

Since the Data Challenge Aegean has continued development and the latest version can be found on
GitHub\footnote{https://github.com/PaulHancock/Aegean}. The following enhancements and improvements
have been made which would improve the performance of Aegean in this data challenge, were it to be run again:
\begin{itemize}
\item The Background And Noise Estimation tool (BANE, also available on GitHub) can provide more accurate
background and noise images than those created by Aegean. The use of these images has been shown to increase
the reliability and flux accuracy of Aegean on other real world data sets.
\item Aegean now produces more accurate source shape and position angle measurements for all WCS projections
\item A bug that cased a small systematic offset in RA/DEC has been fixed. The offset was of the order of one pixel.
\item In the Data Challenge Aegean was instructed not to fit islands with more than 5 components. Islands with more
than 5 components were reported with their initial parameters instead of their fit parameters. The current version of
Aegean is now able to fit the brightest 5 components and estimate the remainder. This may improve the accuracy of
the flux density for bright sources that are in an island with many components.
\end{itemize}

PySE was developed as a component of the LOFAR Transients Pipeline\footnote{http://docs.transientskp.org/}
\citep[or ``TraP''; ][]{Swi:15},
which provides an end-to-end system for detecting and characterising transients
in an image stream. Since the work described in this paper, the TraP, including PySE, has been released as an
open source project under a BSD-style license. It is available for download from
GitHub\footnote{https://github.com/transientskp/tkp} and contributions from the community are welcomed.
Since 2013 the main addition to PySE has been the option to monitor specific positions in an image stream.
The user, or the pipeline, can specify a position from which PySE will extract flux even if no sources are identified.
This is important when building light curves for transient sources.

Duchamp's shortcomings identified by this analysis are expected. The aim of Duchamp is to provide locations of islands of
significant pixels only, and to parameterise the detected islands based solely on the detected pixels, not through fitting of
analytic models. This feature has not (yet) been incorporated into Duchamp, as its focus is primarily on three-dimensional,
spectral-line source-finding. Selavy represents the adaption of the Duchamp software for continuum source-finding and
parameterisation.

Selavy is the prototype ASKAPsoft source-finder that is under development and has been continually refined since the
Data Challenge was run. Development has focused principally on improving the background and noise estimation,
using a sliding box approach to measure the local noise, corresponding to the Selavy (box) mode used here albeit
improved in reliability, and on improving the
determination of the initial conditions for the Gaussian fit. This has benefited from input from the EMU source-finding group, in
particular the approaches used for Aegean described in \citet{Han:12}. These improvements will help the completeness
arising from the Gaussian fitting, in particular for cases where multiple Gaussians are required (see discussion
in \S\,\ref{CandR}). As the ASKAPsoft pipelines evolve through the commissioning of the Boolardy Engineering Test
Array and the full ASKAP telescope we
expect to incorporate further improvements encapsulating lessons learnt from this and any subsequent Data Challenges.

\subsection{ATLAS source finding experience}
\label{atlas}
The Australia Telescope Large Area Survey \citep[ATLAS,][Banfield et al., in prep]{Nor:06, Mid:08, Hal:14a, Hal:14b, Fra:15} is a
survey of 6.3 square degrees with a resolution and sensitivity similar to those of EMU, and is being used as a testbed for EMU.
Source extraction for Data Release 2 (DR2) of ATLAS was performed using a combination of {\sc blobcat} and IMFIT, the latter
as part of a semi-automated pipeline for following up blended sources that were flagged by {\sc blobcat}, as described by
\citet{Hal:14a,Hal:14b}.
The Data Challenge described in this paper was completed before source extraction of the final ATLAS Data Release 3 
\citep[DR3;][Banfield et al., in prep]{Fra:15}, and preliminary versions of the results presented here were used to inform the
ATLAS source extraction. For the ATLAS DR3 source extraction the four finders {\sc blobcat}, Aegean, PyBDSM and
SOURCE\_FIND were tested. The differences were found to be small between these finders, and ultimately
{\sc blobcat} was used because it takes bandwidth smearing and peak bias into account. Complex sources
identified by {\sc blobcat} were fit with multiple Gaussians using the task IMFIT. Ambiguity over the number of
Gaussian components to fit sometimes led to the necessity of a post-processing step to merge nearby Gaussians,
which in turn led to the question of when two components should be merged. The criterion was adopted that
two Gaussian components were merged if the flux density distribution did not show a significant minimum between
the two components. This may be related to the effect seen in the context of the catastrophic flux density estimates
discussed in \S\,\ref{fluxes} above. It is also worth noting that 0.7\% of the ATLAS sources were manually identified as
spurious and removed from the catalogue. All such spurious sources corresponded to image artifacts close to the
brightest sources. This real-world experience demonstrates
that we do not yet have an automated source finder suitable for large surveys, but that further development of the best
existing finders is necessary.

\section{CONCLUSIONS}
\label{conc}

We have presented the ASKAP/EMU Data Challenge, and described the result of assessing the performance of source finding
tools on simulated images, to establish the strengths and limitations of existing automated source finding and
measurement approaches. Three Challenge images were used, presenting a range of different source distributions
and properties. Nine teams participated, with eleven source finders being tested. Our analysis explores the completeness
and reliability of the submitted finders, their ability to reproduce the image flux distribution, and their performance
in characterising the position and flux densities of the measured sources.

One limitation of the current Data Challenge was the broad scope of the analysis attempted, even when limited primarily
to point sources. During the analysis it became clear that there are a large number of areas that would benefit from focused
investigation, in particular those related to the detection and characterisation of overlapping or blended sources,
and complex source structures, as well as to catastrophic outliers, and subtle but systematic effects in
the estimation of source flux densities. Future Data Challenges may choose to focus explicitly on
a more narrow range of performance areas in order to allow themselves the scope to investigate the details
more deeply than has been possible in the existing investigation. There were also practical limitations to the current
Challenge images, such as the sources being assigned to pixel centres, that should be relaxed and explored
in detail in future work.

The various finders that were blindly applied to the Challenge images produce completeness and reliability levels at
or close to 100\% at sensitivities above $\approx 10\,\sigma$, and declining much as expected at fainter sensitivities.
Each tested finder exhibits limitations to a greater or lesser degree. While no finder performed best across all the
tested elements, those that performed well include Aegean, APEX, {\sc blobcat}, IFCA, PyBDSM (sources), PySE and
SOURCE\_FIND. SExtractor performed more poorly than most other finders in terms of flux density estimation, although
demonstrating reasonable completeness and reliability. The other tested finders showed limitations to some degree
in either completeness, reliability or flux density estimation.

We also tested Duchamp and Selavy, finders both authored by Whiting, one of the Challenge initiators.
Duchamp, originally designed for identifying neutral hydrogen emission in radio data cubes, was not expected to perform well
in this analysis for a variety of well-known reasons, and was included for completeness. Selavy was tested as it is the current
implementation of the ASKAPsoft source finder, and provides an important assessment of the likely current performance of the
ASKAPsoft pipeline measurements.

Clear outcomes have been established in terms of identifying areas to improve, both for Selavy and the ASKAPsoft
source finder, as well as the other tested finders individually. It is obvious that accurate characterisation of completeness
and reliability is a requirement in order to have accurate statistical constraints on the performance of any finder. 
The positional accuracy of measured point sources is generally good in almost all finders, but here CuTEx
performed better than the others, suggesting that its fitting approach has an advantage in minimising the rms of fitted
positional uncertainties. In terms of flux density estimation, APEX, {\sc blobcat}, PyBDSM (sources), PySE and
SOURCE\_FIND in particular perform well, with well-constrained
uncertainties and minimal outliers. The fraction of catastrophic outliers in flux density estimation, at best around $0.2\%$ from
all tested finders, will need to be reduced to ensure high fidelity performance for future sky surveys.

Here we summarise the key outcomes that would benefit ASKAPsoft and future source finder
development, with an indication of which of the tested finders may provide suitable algorithms or approaches:
\begin{itemize}
\item Quantifying completeness and reliability accurately as a function of S/N through repeated simulations and testing.
\item Robust handling of blended sources (this affects completeness, reliability and flux density estimation,
 see \SS\,\ref{CandR} and \ref{fluxes}). Aegean and {\sc blobcat} are examples using different approaches that
 each work well in this regime.
\item Source position estimation (this is already good in all finders, we are looking to capitalise on the best performance,
 see \S\,\ref{positions}). CuTEx, PySE and PyBDSM demonstrated the best performance in this aspect for the
 current investigation.
\item Identifying the origin of and rectifying the flux density overestimates at faint levels, as seen in Selavy (\S\,\ref{fluxes}).
 Finders that did not show this effect include APEX, {\sc blobcat}, CuTEx, IFCA, PySE, SAD and SOURCE\_FIND.
\item Identifying the origin of and minimising (or eliminating) the fraction of catastrophic outliers (\S\,\ref{fluxes}).
 Finders with the lowest such fractions currently include Aegean, APEX, PyBDSM (sources), PySE, and SOURCE\_FIND.
\item Capitalising on the strong performance of IFCA in accurately measuring flux densities to very low S/N
(\S\,\ref{fluxes}).
\item Robustly detecting and characterising extended or complex sources (\S\,\ref{CandR} and \ref{iman}).
 This is a challenging area to quantify even for simple extended Gaussian sources in the presence of neighbouring and
 blended sources. Effort is needed to accurately quantify the performance of finders here more extensively than has been
 attempted in the current analysis. The performance of the different modes of Selavy in the image-based analysis
 (\S\,\ref{iman}, for example, suggest that some complex combination of its detection and characterisation stages in
 the different modes, informed by the performance of other finders, may be worth implementing. Within the limitations
 imposed by the current analysis, finders that perform well in this area include
 {\sc blobcat}, PyBDSM, PySE (D5A3), and SExtractor.
\item Automating the still manual process of identifying and excluding or flagging imaging artifacts (\S\,\ref{atlas}).
\end{itemize}

The most successful approaches for each of these elements will need to be combined in order to implement
the most robust approach to source finding for future generations of high-sensitivity all-sky radio surveys.

\begin{acknowledgements}
We thank the referee for suggestions that have helped to improve the accuracy and presentation
of this work.
This work was supported by resources provided by the Pawsey Supercomputing Centre with funding from
the Australian Government and the Government of Western Australia.
NS is the recipient of an ARC Future Fellowship.
The LOFAR team acknowledges Hanno Spreeuw as the original developer of PySE.
The IFCA team acknowledge Spanish MINECO Projects AYA2012-39475-C02-01 and
Consolider-Ingenio 2010 CSD2010-00064.
RPB has received funding from the European Union Seventh Framework Programme under grant agreement
PIIF-GA-2012-332393.
CF acknowledges financial support the {\it ``Agence Nationale de la Recherche"} through grant
ANR-09-JCJC-0001-01 and ANR-14-CE23-0004-01.
HG acknowledges financial support from the UnivEarthS Labex program of Sorbonne
Paris Cit{\'e} (ANR-10-LABX-0023 and ANR-11-IDEX-0005-02).
PH has been supported by the Australian Research Council through the Super Science Fellowship grant FS100100033.
SM and MP acknowledge funding from Contracts I/038/080/0 and I/029/12/0 from Agenzia Spaziale Italiana.
JGN acknowledges financial support from the Spanish MINECO for a ``Ramon y Cajal" fellowship,
cofunded by the European Social Fund.
LR acknowledges support from the US National Science Foundation under grant AST12-11595 to the University of Minnesota.
ES acknowledges support, at the date of this work, from the NASA Astrophysics Data Analysis Program grant NNX12AE18G.
Parts of this research were supported by the Australian Research Council Centre of Excellence for All-sky Astrophysics
(CAASTRO), under grant number CE110001020.
The National Radio Astronomy Observatory is a facility of the National Science Foundation operated under cooperative agreement by Associated Universities, Inc.
\end{acknowledgements}


\appendix
\section{DESCRIPTION OF SOURCE FINDERS}
\label{app1}

For ease of reference we provide here descriptions of the finders submitted for the Data Challenge
describing their methods of operation and different modes of use if applicable.

\subsection{Aegean}
Aegean has been designed to find and characterise compact sources in radio images. The underlying algorithms
are built with the assumption that the user is interested in objects that can be well characterised by a number of
Gaussian components. This focus on compact sources means that Aegean will produce a rather complex
characterisation of extended sources or resolved structures, which will be of limited use. The current version of
Aegean has an alternate mode of operation which provides a characterisation scheme that is more appropriate
for amorphous or resolved structures. This alternate mode of operation characterises a single island as a
single ``blob" in much the same way that {\sc blobcat} does.

In this data challenge Aegean r808\footnote{http://www.physics.usyd.edu.au/$\sim$hancock/files/Aegean.808.tar}
was used. Aegean identifies significant pixels (finds sources) by calculating a noise image from the interquartile range of
pixels in regions of size $30\times30$ synthesised beams, forming an image that represents signal-to-noise,
and finally selecting all pixels above a given threshold. In this challenge a threshold of $5\,\sigma$ was used. Once
significant pixels are identified, a flood-fill algorithm is run to group these pixels together into islands, and
the islands are expanded to include adjoining pixels that are have S/N$\ge 4$. This means
that islands of pixels are seeded with a threshold of 5 and grown with a threshold of 4. The Aegean source
characterisation stage operates on one island at a time, and involves the creation of a curvature map.
The curvature map represents the second derivative of the input image, and is negative at and around
local maxima. To determine how many components are within an island Aegean counts the number of local
maxima within the island, each local maximum is assigned a single component. Islands are thus fit with
multiple Gaussian components. The fit is achieved using a constrained least squares
Levenberg-Marquardt algorithm. The position, flux, and shape of each component is constrained to
prevent them from merging with each other, and to avoid unphysical results.

Aegean can be downloaded from the Astrophysics Source Code
Library\footnote{http://ascl.net/phpBB3/viewtopic.php?t=30381}.

\subsection{APEX}

Astronomical Point source EXtractor
\citep[APEX\footnote{http://irsa.ipac.caltech.edu/data/SPITZER/docs/dataanalysistools/tools/mopex/},][]{MM:05}
is the source extraction program included in the
Mosaicking and Point-source Extraction (MOPEX) package that was developed for {\em Spitzer\/} Space Telescope data.
APEX is similar to other thresholding source extraction algorithms in that it performs background and noise estimation,
but detected clusters of pixels are fitted with a point response function (PRF) to return fitted point sources. APEX allows
both passive and active deblending to handle crowded fields. In passive deblending the detected point sources are
determined to be in close proximity such that their PRFs overlap, and APEX then fits them simultaneously. Active
deblending is where a single point source fit fails and APEX then fits the cluster of pixels with multiple point sources.
APEX also allows the user to specify an arbitrary number of apertures for aperture photometry.
APEX from MOPEX v18.5 was used for this Challenge.

\subsection{Blobcat}

{\sc blobcat}\footnote{http://blobcat.sourceforge.net/} is described by \citet{Hal:12}.
{\sc blobcat} is designed to operate not only on images of total intensity but also linear polarization. Version 1.0 was
used for this Data Challenge. Due to the Challenge's focus on point-like sources, no effort was made to decompose blobs
that were flagged by {\sc blobcat} as likely consisting of blended sources. This should be considered when interpreting results
in this paper. For an example application where blended sources are accounted for in a semi-automated pipeline with
MIRIAD's IMFIT algorithm, see analysis of ATLAS DR2 by \citet{Hal:14a,Hal:14b}. Suggestions for improving {\sc blobcat}
are always welcome; please see the web link for contact details. 

\subsection{CuTEx}

CuTEx \citep[Curvature Threshold Extractor,][]{Mol:11} is an IDL-package that was developed (and is extensively used)
within the framework of the Open
Time Key Project on the {\em Herschel\/} satellite called Hi-GAL \citep{Mol:10}. This program gathered data in 5 bands
(70, 160, 250, 350, 500 $\mu$m) of the entire Galactic plane, with the aim of studying the early stages of the
formation of (high-mass) stars across the Galaxy. CuTEx was designed to enhance compact sources (sizes not larger
than 3 times the instrumental point spread function) in the presence of an intense and highly variable background such
as that seen in {\em Herschel\/} observations of the Galactic Plane. 
The CuTEx package is divided into two parts, a detection element that identifies sources and a photometry extraction
element that measures their sizes and fluxes.
Compact sources are detected by analysing the second derivative of the images in four directions,
which is proportional to the ``curvature" of the intensity. In those derivative images, all large scale emission is
damped (in the case of infrared images it is the background), while all peaked objects (compact sources) are enhanced.
Candidate sources are identified by associating contiguous pixels with a value of the second derivative in excess of a
certain threshold and grouped into small clusters. Clusters can contain more than one source, in which case they will
be extracted as a group. At this stage an estimate of the sizes of the sources is also performed by measuring the
distance between two opposite ``first most negative" values of the second derivative around the identified centre
of the source (there will be eight points) and fitting these with an ellipse, with the aim to obtain an initial guess for the
photometry extraction. The photometry extraction part uses this list of candidates to determine the integrated flux
and the background values on the original image (in our case the restored image) by fitting elliptical Gaussians, and
measures the peak flux as well as the FWHM in two orthogonal directions and position angle (PA) of the fitted Gaussian.
The fitting engine used is the Markwardt MPFIT package and strong constrains on the large number of parameters for
each sources are applied to ensure convergence of the fit.

\subsection{IFCA}
The IFCA source finding approach used in this challenge is a combination of SExtractor and optimal filtering kernels.
Two methods were used, referred to as IFCA (MF) and IFCA (BAF).

IFCA (MF) is so named as it uses matched filters.
The matched filter kernels have been obtained iteratively for each one of the three Challenge images as follows. In
each iteration we estimate the power spectrum of the background fluctuations. This power spectrum is
used to calculate the optimal matched filter. The image is then filtered and all sources above the $4\,\sigma$
level are detected and subtracted from the image. The new image with the sources above the $4\,\sigma$
level is used as input for the next iteration until convergence is achieved (no new $4\,\sigma$ detections
arise). The rms of the final filtered images (with all the $5\,\sigma$ detections subtracted) are the estimates
of the backgrounds we use to decide the detection threshold for our catalogues. Some details of the
matched filter used here can be found in \citet{Lop:06} and references therein.

IFCA (BAF) is so named as it uses a biparametric adaptive filter.
The biparametric adaptive filter \citep{Lop:12} kernel has been obtained as follows for each of the three Challenge
images. We iteratively explored the two-parameter space that defines our filter (the index of the filter $n$,
that is related to the index of the power-law that best describes the statistical properties of the background
of the images; and the scale of the filter $R$) to look for a minimum in the rms of the filtered field. For
Challenges~1 and 2 we used a kernel with $n=0$ and $R=0.65$, whereas for Challenge~3 we
used $n=2$ and $R=0.5$. The reason for using a filter with a higher index $n$ in Challenge~3 is because
of the presence of extended objects (local galaxies or galaxy-like structures). Since this Challenge is
devoted to point-source detection and extraction, this particular kernel is able to easily remove structures
in the images that are very different from the point spread function, as in the case of local extended
galaxies, before attempting to do any detection. The rms of the final filtered image is obtained after
masking all the detections above ${\rm S/N}>4$ in the image. As in the previous case, three different
estimates of the rms have been calculated and used to set a S/N cut in the catalogues. The process
of iteratively finding the the parameters that are used to build the kernel is quick and can be easily
automated. For all-sky Healpix fits images this code exists and is automatic. For this Challenge things
have been done in a partially automated fashion as this was the first time we applied such a filter to images other
than cosmic microwave background or sub-millimetre images.

The details of the biparametric adaptive filter can be found in \citet{Lop:12}.
An additional reference of interest, since the IFCA-BAF filter under some circumstances
defaults to the Mexican Hat Wavelet family, can be found in \citet{Gon:06}.

\subsection{PyBDSM}

PyBDSM\footnote{https://dl.dropboxusercontent.com/u/1948170/html/index.html}
\citep[``Python Blob Detection and Source Measurement'', a Python source-finding software
package written by Niruj Mohan Ramanujam, Alexander Usov and David Rafferty;][]{MR:15} calculates
rms and mean images and then identifies islands of contiguous significant emission, computed
either by a hard threshold or by using the False Detection Rate algorithm \citep{Hop:02}. PyBDSM allows fitting
of one or multiple Gaussians to each island and grouping of nearby Gaussians within an island into
``physical" sources. A modified fitting routine can also handle extremely extended sources.
It can also decompose islands into shapelet coefficients. In addition a PyBDSM module is
available to decompose the residual image resulting from the normal fitting of Gaussians into
wavelet images of various scales, and building these back into sources using the pyramidal morphological
transform. This step is useful for automatic detection of diffuse sources. Errors on each of the fitted
source parameters are computed using the formulae in \citet{Con:97}. PyBDSM can also
calculate the variation of the point spread function across the image using shapelets, and
calculate the spectral index of sources. 

In this work we define an island threshold at 3$\sigma$ to determine
the region to which source fitting is done and an additional limit
parameter at 5$\sigma$ in such a way that only islands with peaks
above this absolute threshold will be used. In addition, we have taken
into account both the catalogue containing all the fitted Gaussians, referred
to as ``PyBDSM (Gaussians)," and the catalogue in which Gaussians
have been grouped into sources, referred to as ``PyBDSM (Sources)."

\subsection{PySE}

PySE\footnote{http://docs.transientskp.org/tkp/r2.0.0/tools/pyse.html}
was developed within the LOFAR Transients Key Science Project
\citep{vanH:13,Fen:07} as part of its real-time transient search pipeline. On the assumption that
(relatively) fast radio transients are unresolved, the software is optimised
for the detection of point-like sources.
PySE processing fundamentally involves the following steps:
\begin{enumerate}
  \item{The image is divided into rectangular cells, and the pixel values in
  each cell are iteratively $\sigma$-clipped around the median;}
  \item{Bilinear interpolation of the mean across cells is used to derive a
  background map, which is subtracted from the data;}
  \item{Bilinear interpolation of the standard deviation across cells is used
  to calculate an rms noise map;}
  \item{Groups of contiguous pixels at some detection threshold over the rms
  noise are selected as potential source peaks;}
  \item{Pixel groups are extended to include surrounding pixels above some
  (lower) analysis threshold;}
  \item{Optionally, pixel groups are decomposed (or ``deblended'') into their
  constituent parts where applicable;}
  \item{Source properties are estimated by means of a least-squares fit of an
  elliptical Gaussian.}
\end{enumerate}
User configuration is required to select an appropriate cell size: smaller
cells are better suited to tracking variation across the image, but are more
sensitive to bias from bright sources.
The detection and analysis thresholds may be specified directly by the user,
or alternatively can be derived using a False Detection Rate 
algorithm \citep{Hop:02}.
Some source properties may be held constrained during fitting, in particular,
when measuring unresolved sources, it may be appropriate to constrain the
source shape to be equal to that of the restoring beam.

For this analysis we used an unreleased prototype of PySE from late 2012.
Two catalogs were provided both for Challenge~1 and for
Challenge~2 (``PySE (D5A3)'' and ``PySE (FDR)'') and one catalogue for
Challenge~3 (``PySE (D5A3)''). ``D5A3'' refers to detection and analysis
thresholds of $5\,\sigma$ and $3\,\sigma$ respectively, while ``FDR'' is configured to use
the False Detection Rate algorithm with a 1\% error rate.
Square cells of side 50 pixels were used for calculating the background and
noise maps in Challenges~1 and 2; 30 pixel squares were used for Challenge~3.
In each case we used the option to constrain the shape of the extracted sources
to be equal to the restoring beam and to decompose sources lying within the same island;
all the other options were left to their default values.

A detailed description of the algorithm may be found in 
\citet{Spr:10}, \citet{Swi:15}, and \citet{Car:14}.

\subsection{SAD}
SAD (Search and Destroy) is an automated source finding algorithm implemented within the Astronomical Image
Processing System (AIPS). It was developed to create the source catalogue for the NRAO VLA Sky Survey project
\citep{Con:98}. Sources in the image are fit with 2D Gaussian functions.  The strongest source is fit and then
removed (i.e.\ searched and destroyed), and the process repeated until a stopping threshold is reached ({\tt CPARM}).
SAD can fit a maximum of 40,000 sources per run, so we split the Challenge~1 and 2 images into two east and west
sections and fit these independently. Challenge~3 was processed as a single field, but at two resolutions.
The Challenge~3 image was blanked to mask extended sources. The image was then searched for sources at
the full resolution (highres). We then restored the blanked regions in the residual image, convolved it to
$30''$ resolution and searched the resulting image (lowres).

The SAD stopping threshold was set to 0.04, 0.004, 0.01 and 0.025 Jy~beam$^{-1}$ respectively for
Challenges~1, 2, 3 highres and 3 lowres. In addition several criteria were applied to reject sources based on the
parameters of the fitted solutions. These are set using inputs {\tt DPARMS}, which reject based on peak
and total flux, source width and location of the peak relative to the fitted region (island).
Peak and total flux rejection criteria were set to below the stopping threshold. Fits with very large
widths were rejected as were fits with peak positions outside of the island. If the rms of the residual to
a single component Gaussian fit is above a threshold ({\tt ICUT}), then multiple Gaussians of increasing
number are fitted simultaneously. {\tt ICUT} was set equal to the stopping threshold values for each run.

Python scripts were written to merge the sources from the split images of Challenges~1 and 2, and the two
resolutions of Challenge~3. As a final check the Gaussian peak for each fit was checked against the image
value at fitted peak position. If the image data value was less than 30\% of the fitted peak the sources was
considered spurious and removed. 

\subsection{SExtractor}

SExtractor\footnote{http://www.astromatic.net/software/sextractor}
is a tool commonly used with optical astronomy images to perform automated detection and photometry of
sources \citep{Ber:96}. It is oriented towards the reduction of large surveys of galaxies, but can also perform well
in moderately crowded star fields. Analysis of the astronomical image is done in two passes. The first pass builds
a model of the sky background and calculates global statistics. During the second pass the image is optionally
background-subtracted and filtered. SExtractor uses a threshold technique to isolate groups of pixels as detected
``islands". These are then deblended and measured for source size, position and flux. SExtractor v2.8.6 was
used for this Challenge, with two smoothing scales for estimating the background sky model, corresponding to
10 or 30 times the resolution element or point spread function, referred to as SExtractor (10 beam) and
SExtractor (30 beam) respectively. This choice was
informed by previous analysis of SExtractor's performance on radio images \citep{Huy:12}.

\subsection{SOURCE\_FIND}
The SOURCE\_FIND software is described in detail in \citet{Fra:11}, where it is applied to the 10C survey of
radio sources at 16 GHz. The software is capable of identifying and characterising sources in radio synthesis maps
with varying noise levels and synthesised beams, and includes a straightforward and accurate method for distinguishing
between point-like and extended sources over a wide range of SNRs. It is part of the standard data reduction pipeline
for the Arcminute Microkelvin Imager \citep{Zwa:08}.

The first step in the source extraction process involves determination of the noise level. At each pixel position in the
image the noise is taken as the rms inside a square centred on the pixel whose width is set to some multiple of the
synthesised beam and, in order to avoid the noise estimate from being significantly affected by source emission, points are
clipped iteratively until convergence at $\pm3\,\sigma$ is reached. The width of the sliding box for noise estimation
was set to 20 times the synthesised beam size for Challenges~1 and 2, and to 40 times the synthesised beam
for Challenge~3.

The noise map is used to identify sources on the basis of their S/N. In all three data challenges, local maxima above
$5\,\sigma$ were identified as sources. A peak position and flux density are measured by interpolating between the
grid points. This is done by calculating the map values on a successively finer grid (up to 128 times finer), by repeated
convolution with a Gaussian-graded sinc function \citep{Ree:90}. Here we did not use the Gaussian fitting mode of
SOURCE\_FIND to measure integrated flux densities, centroid positions and source sizes. Rather, these parameters
were measured by integrating contiguous pixels down to a lowest contour level of $2.5\,\sigma$.

\subsection{Finders tested by the Challenge organisers: Duchamp and Selavy}

Duchamp\footnote{http://www.atnf.csiro.au/people/Matthew.Whiting/Duchamp} is a source-finder designed to find
and describe sources in three-dimensional spectral-line data cubes \citep{Whi:12}, but is readily applied to
two-dimensional images. The source-detection performed by Duchamp is based on simple flux or S/N
thresholding, with an optional secondary threshold to which detected sources are grown (to increase their size
and reliability). The detectability of sources is enhanced by using one of several pre-processing methods
that aim to reduce the noise yet preserve astronomically-interesting structures in the data.
One pre-processing method is to smooth the data with a defined kernel, either spatially or spectrally, and
then perform the search on the smoothed data. The alternative pre-processing method is to use the {\`a} trous wavelet
algorithm to generate a multi-resolution wavelet set, showing the amount of signal as a function of scale size and
position in the data set. Each wavelet array (i.e. corresponding to a single scale size) has a threshold applied,
and pixels with values below this threshold set to zero. The thresholded wavelet arrays are then added back together
to provide an array that has a large fraction of the noise removed. A worked example in one dimension is given in
\citet{Whi:12}. Duchamp provides a parameterisation of the detected sources, calculating values such as integrated flux,
principle axes and weighted centroid position based only on the detected pixels. Duchamp is intended to act as a tool for
providing the location of interesting features, yet remain agnostic as to their intrinsic shape, and so provides no source
fitting (such as the Gaussian fitting typically used in continuum image analysis). This approach, however, does
lead to the characteristic error pattern seen in Figure\,\ref{flxcmp1}.

For this Data Challenge, we used version 1.2.2 of Duchamp to generate results with three distinct modes:
Duchamp (basic) used simple signal-to-noise threshold without pre-processing; Duchamp (smooth) used a
6-pixel FWHM 2D Gaussian kernel to smooth the data prior to searching; and Duchamp ({\`a} trous)
used a 2-dimensional {\`a} trous algorithm with a $4\,\sigma$ wavelet threshold to reconstruct the noiseless
data prior to searching.

Selavy is the prototype ASKAP pipeline source-finder \citep{WH:12} that is being developed as part of the
ASKAP Science Data Processing software (also known as ASKAPsoft). Selavy builds on the Duchamp software library,
providing additional functionality that is necessary to run in a high-performance pipeline environment on a range of
image types, most notably 2D Gaussian fitting to detected sources, a spatially-variable threshold that responds to
local noise, and the ability to run in parallel on a high-performance supercomputer. Duchamp assumes a single threshold
for the entire dataset, which gives a uniform selection criterion, but can have drawbacks in the presence of non-uniform noise.
Selavy overcomes this in one of two ways. First, it can remove the large scale variation brought about by primary beam
effects by dividing through by a weight image. Searching is then performed on the de-weighted image, but parameterisation
is still done on the original (where the fluxes should be correct). The second way is to find the local noise at each pixel, by
measuring it within a local box region. This allows a signal-to-noise threshold to rise where there is strong local noise (for
instance, there may be deconvolution sidelobes around a bright source) and decrease where the noise is low.
The Gaussian fitting takes a given Duchamp detection (an island) and fits a number of 2D Gaussian components to the
pixels in that island. The number of components to fit, and the initial estimates of their parameters, are determined by
applying a large number of sub-thresholds to the island, ranging from the detection threshold to the peak. This approach
works well when enough sub-thresholds are applied, but too few may result in secondary components being missed
(which may be the case in some situations in this Data Challenge, e.g., \S\,\ref{CandR}). More recent versions of Selavy
have incorporated the curvature-map method of determining local maxima use by Aegean, which is proving to be successful.
For this Data Challenge Selavy was run using v1.2.2 of the Duchamp library, in the same modes as Duchamp, plus two
additions: Selavy (weight) used the weights image to scale the noise across the field prior to searching; while Selavy (box) used a $101\times 101$ box to find the local noise prior to searching with a S/N threshold.

\end{document}